\documentclass[lettersize,journal]{IEEEtran}
\usepackage{amsmath,amsfonts}
 \usepackage{makecell}
\usepackage{algorithm}
\usepackage{array}
\usepackage[caption=false,font=normalsize,labelfont=sf,textfont=sf]{subfig}
\usepackage{textcomp}
\usepackage{stfloats}
\usepackage{url}
\usepackage{verbatim}
\usepackage{graphicx}
\usepackage{cite}
\usepackage{amssymb}
\usepackage{algpseudocode}
\usepackage{hyperref}
\usepackage{booktabs}
\newcommand{\method}{LEAN-3D}
\include{reference}
\begin{document}

\title{\method{}: Low-latency Hierarchical Point Cloud Codec\\ for Mobile 3D Streaming}
\author{Yuchen Gao,~\IEEEmembership{Student Member,~IEEE},
        and Qi Zhang,~\IEEEmembership{Senior Member,~IEEE}%
\thanks{Yuchen Gao and Qi Zhang are with the Department of Electrical and Computer Engineering, Aarhus University, Aarhus, Denmark (e-mail: yuchen@ece.au.dk; qz@ece.au.dk).}%

\thanks{This research was supported by the TOAST project, funded by the European Union’s Horizon Europe research and innovation program under the Marie Skłodowska-Curie Actions Doctoral Network (Grant Agreement No. 101073465) and NordForsk Nordic University Cooperation on Edge Intelligence (Grant No. 168043).} 

}


 
\maketitle
\begin{abstract}
We aim to make learned point cloud compression deployable for low-latency streaming on mobile systems. 
While learned point cloud compression has shown strong coding efficiency, practical deployment on mobile platforms remains challenging because neural inference and entropy coding still incur substantial runtime overhead. This issue is critical for immersive 3D communication, where dense geometry must be delivered under tight end-to-end (E2E) latency and compute constraints.

In this paper, we present \method{}, a compute-aware point cloud codec for low-latency streaming. \method{} designs a lightweight learned occupancy model at the shallow levels of a sparse occupancy hierarchy, where structural uncertainty is highest, and develops a lightweight deterministic coding scheme for the deep hierarchy tailored to the near-unary regime. We implement the complete encoder/decoder pipeline and evaluate it on an NVIDIA Jetson Orin Nano edge device and a desktop host. In addition, \method{} addresses the decoding failures observed in cross-platform deployment of learned codecs. Such failures arise from numerical inconsistencies in lossless entropy decoding across heterogeneous platforms. Experiments show that \method{} achieves 3-5$\times$ latency reduction across datasets, reduces total edge-side energy consumption by up to 5.1$\times$, and delivers lower sustained E2E latency under bandwidth-limited streaming. These results bring learned point cloud compression closer to deployable mobile 3D streaming.
 
\end{abstract}
 
\section{Introduction}
\begin{figure}[t]
  \centering
  \includegraphics[width=\linewidth,keepaspectratio, trim=.4cm 0.3cm 1.3cm 0.3cm, clip]{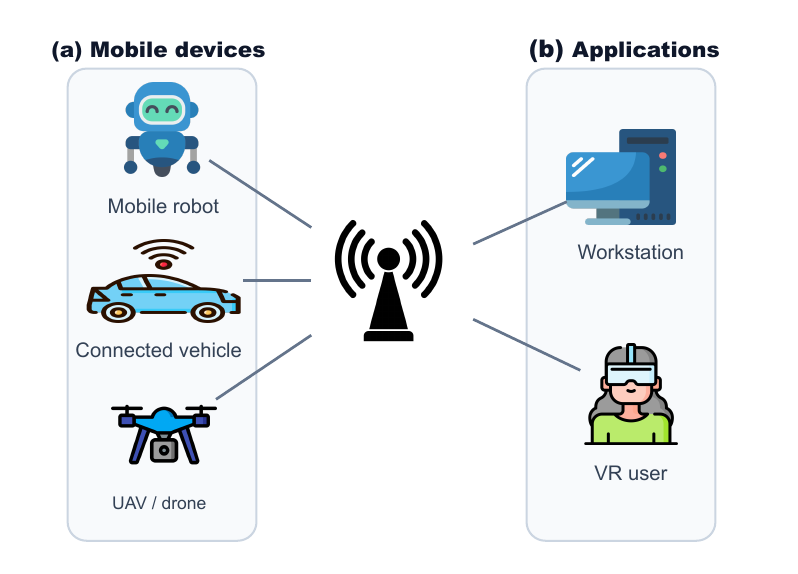}
  \caption{Deployment scenario of \method{}. (a) Portable sensing devices acquire dense point clouds on a resource-constrained mobile platform. (b) Remote hosts for visualization or downstream processing.}
  \label{fig:usecase}
\end{figure}

\IEEEPARstart{P}{oint} clouds captured by LiDAR and RGB-D sensors have become a fundamental 3D representation for robotics, autonomous driving, digital twins, and immersive telepresence. In mobile systems such as robots, drones, and handheld scanners, dense 3D measurements are produced continuously and often need to be streamed to a remote operator, a nearby host, or a cloud service for mapping, perception, and decision-making. However, point clouds are intrinsically data-heavy: practical streams can contain hundreds of thousands to millions of points per frame, and naive transmission quickly becomes bandwidth-prohibitive~\cite{schwarz2019emerging,graziosi2020overview,cao2019survey}. More importantly, in mobile settings practical performance is determined not only by bitrate, but also by on-device codec execution, memory access operations, and energy consumption. These constraints motivate point cloud compression (PCC) methods that are not only compact, but also deployable on resource-constrained mobile hardware~\cite{itu2014tactile,fettweis2014tactile,3gpp_tr38913}. Fig.~\ref{fig:usecase} illustrates the target deployment scenario considered in this work, where a portable or mobile sensing device continuously acquires point clouds and streams compressed geometry to a remote host for visualization or downstream processing.
\begin{figure*}[t]
  \centering
  \includegraphics[width=\linewidth,trim=.2cm 0.4cm .2cm 0.2cm, clip]{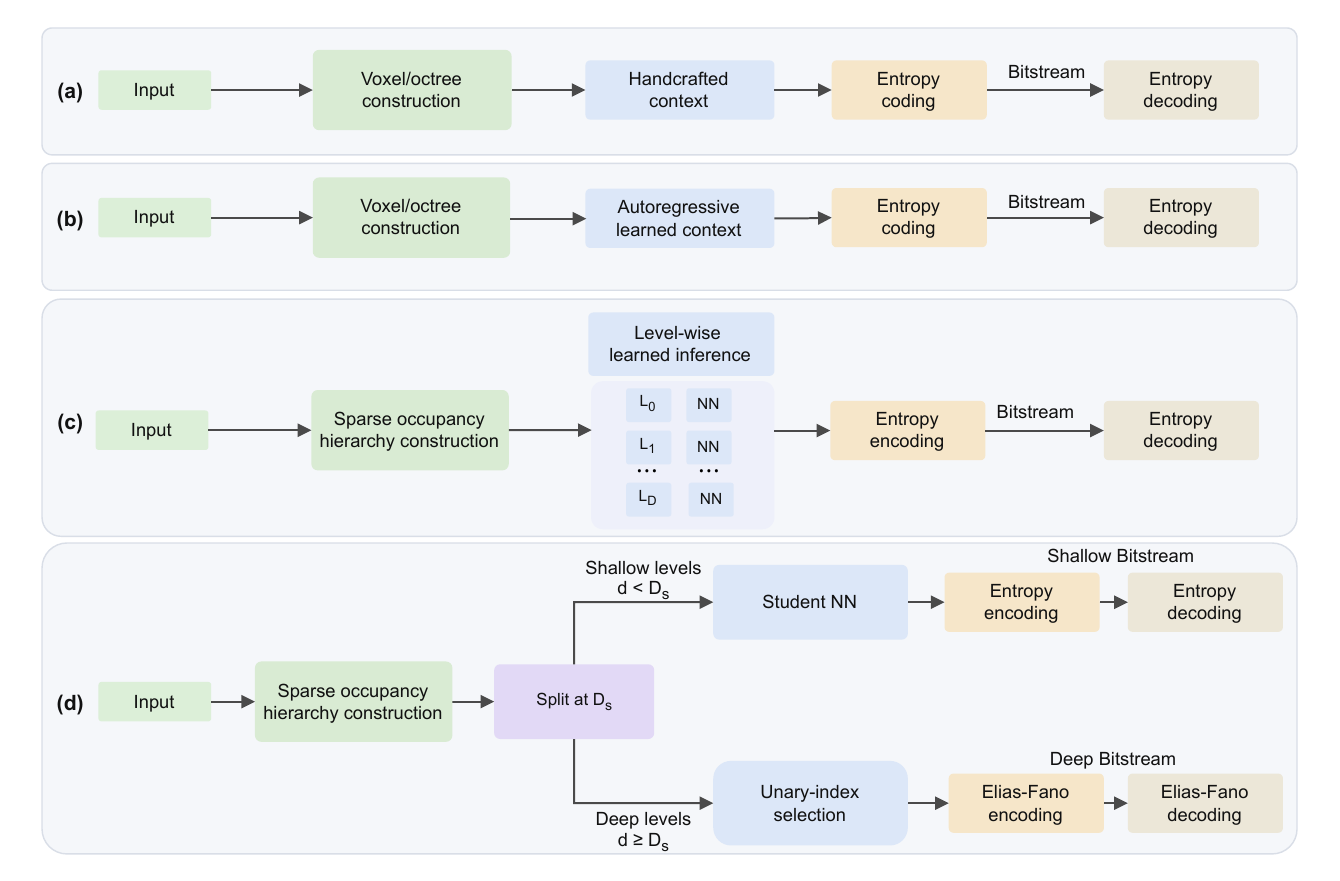}
  \caption{Comparison of four geometry-coding paradigms for point cloud compression. \textbf{(a)} Conventional codecs rely on voxel/octree construction, handcrafted context modeling, and entropy coding. \textbf{(b)} Learned codecs replace handcrafted statistics with autoregressive learned context modeling, while still following a repetitive learned entropy-coding path. \textbf{(c)} RENO-style real-time learned coding organizes geometry as a sparse occupancy hierarchy and applies level-wise batched neural inference across the hierarchy before entropy coding. \textbf{(d)} \method{} adopts a dual-path, compute-aware architecture: shallow levels are assigned to a lightweight student network and entropy-coded into the shallow bitstream, while deep levels are encoded by a deterministic scheme and packed into the deep bitstream using Elias--Fano coding.
This comparison highlights that \method{} is not simply a reduced learned codec, but a hybrid shallow--deep design for lower runtime and more feasible deployment on mobile systems.}
  \label{fig:method_compare}
\end{figure*}

Classical point cloud compression pipelines provide strong coding baselines, but they do not always offer the most favorable rate--runtime operating point for mobile 3D streaming. Compared with recent learned approaches, classical codecs may be less competitive in coding efficiency for some complex geometric distributions, although they remain strong practical baselines in interoperability and engineering maturity~\cite{quach2022survey,gao2025_tpami_survey}. Moreover, their codec pipelines typically rely on multi-stage geometry processing and entropy-coding procedures whose runtime and memory behavior can become difficult to execute efficiently on resource-constrained mobile or portable devices under tight latency and energy budgets~\cite{kammerl2012realtime,cao2019survey}.

Learning-based PCC has substantially improved geometry modeling and coding efficiency by using neural networks to capture 3D structure more effectively. However, a theory-to-practice gap remains for mobile deployment. Prior work is commonly evaluated in terms of compression performance and runtime on desktop-class GPU platforms, whereas in portable systems the central question is whether continuous compression and decompression can be supported under strict compute and energy constraints rather than how closely a method approaches the Rate--Distortion (RD) frontier~\cite{quach2022survey,gao2025_tpami_survey}.

Recent efforts have started to explicitly target real-time learning-based PCC for LiDAR. Notably, RENO proposes a real-time neural codec that compresses multiscale sparse occupancy without explicit octree construction, demonstrating competitive compression with real-time performance on desktop-class GPUs~\cite{you2025reno}. Although this progress demonstrates feasibility, it exposes a deployment mismatch in mobile settings, where the sensing side often relies on Jetson-class or battery-powered compute modules, and the receiving side may also operate on constrained or heterogeneous hardware. In such settings, on-device neural network inference and entropy coding can dominate the available time budget even without considering communication latency. Even with low-queueing network service (e.g., Low Latency, Low Loss, Scalable throughput (L4S)-style approaches), codec runtime can dominate the E2E latency budget~\cite{rfc9330,damigos2025l4s}. Therefore, practical point cloud streaming in mobile settings calls for a codec design that reduces the critical-path runtime while maintaining a competitive compression ratio and robust cross-platform decoding. Fig.~\ref{fig:method_compare} illustrates the data processing pipeline of representative point cloud compression codecs, highlighting the transition from handcrafted methods to fully learned sparse-hierarchy coding and positioning \method{} as a compute-aware hierarchical codec.

Moreover, learned entropy models introduce an additional deployment challenge that has received limited attention in the literature: lossless entropy coding requires the encoder and decoder to construct identical cumulative distribution functions (CDFs). However, floating-point non-associativity and backend-dependent kernels can introduce subtle logit discrepancies across heterogeneous devices. Such discrepancies may change probability quantization boundaries and break entropy synchronization, undermining cross-platform robustness and limiting its extensibility.

 In this paper, we propose \method{}, a compute-aware hierarchical PCC framework for low-latency point cloud streaming on mobile systems, and we implement a working prototype to evaluate its deployment feasibility. 
 
Concretely, \method{} builds on a multiscale occupancy representation in the spirit of real-time learned codecs~\cite{you2025reno}. It combines a distilled predictor for the uncertainty-dominated shallow occupancy hierarchy levels and fast bitwise occupancy construction for deeper levels, rANS coding for occupancy streams at shallow levels~\cite{duda2013ans,giesen2014interleaved}, and Elias--Fano representation for sparse deep-level index sets~\cite{elias1974ef,vigna2013quasi}. To make this design deployable across heterogeneous platforms, we further introduce a bit-exact entropy coding that maps network logits to integer CDFs, thereby preventing cross-platform floating-point discrepancies from breaking entropy synchronization.
 
Our main contributions are as follows:
\begin{itemize}
 \item \textbf{Compute-aware hierarchical geometry codec:}
 We propose a dual-codec architecture that develops learned modeling for shallow levels and deterministic coding for deep levels, achieving approximately 3--5$\times$ lower latency across datasets while preserving learned occupancy modeling where it is most beneficial.

\item \textbf{Bit-exact entropy coding for heterogeneous deployment:}
We introduce a consistent logit-to-CDF construction that guarantees identical integer CDF generation across platforms and eliminates cross-platform decoding failures caused by floating-point discrepancies.

\item \textbf{Working prototype for mobile deployment:}
We implement the full \method{} encoder/decoder pipeline as a working cross-platform prototype for real-world low-latency geometry streaming.

\item \textbf{System-level validation:}
We conduct comprehensive experiments to evaluate \method{}, including full encode/decode runtime analysis, component-wise latency attribution, bandwidth-limited streaming simulation, runtime--rate trade-off analysis, cross-dataset validation, heterogeneous edge-to-host deployment, and edge-side energy measurement.
\end{itemize}

The remainder of this paper is organized as follows: Section~\ref{sec:background} reviews related work on standard and learned PCC. Section~\ref{sec:method} introduces the codec architecture and prototype details. Section~\ref{sec:exp} presents the experimental evaluation. Section~\ref{sec:discussion} discusses the implications and limitations of the current design and outlines directions for future work. Section~\ref{sec:conclusion} concludes the paper.

\section{Related Work and Background}
\label{sec:background}

This section reviews point cloud compression (PCC) and low-latency 3D streaming literature.

\subsection{Standards and Classical Point Cloud Compression}
\label{sec:bg_standards}

Existing PCC standards provide strong baselines for interoperability and RD performance. In particular, MPEG has standardized geometry-based PCC (G-PCC) and video-based point cloud compression (V-PCC) under the ISO/IEC~23090 series~\cite{iso23090_9_gpcc,iso23090_5_v3c}, and the broader MPEG effort has been summarized in influential overviews~\cite{schwarz2019emerging,graziosi2020overview}. Complementary engineering solutions such as Google Draco provide widely deployed geometry compression toolchains~\cite{draco}. Despite their effectiveness, such pipelines are primarily optimized for general-purpose or offline settings and typically involve multi-stage processing, repeated memory traffic, and nontrivial entropy-coding overhead. These costs can become a bottleneck when the codec must run continuously on mobile or portable platforms under strict runtime and energy constraints~\cite{kammerl2012realtime,cao2019survey}.

Classical geometry compression has exploited octree-like hierarchies and predictive coding. Early point-based graphics pipelines already used octree partitioning for compact geometry representation~\cite{schnabel2006octree}. In robotics and online perception systems, real-time point cloud compression has been studied to meet compute and memory constraints in streaming~\cite{kammerl2012realtime}. Although such pipelines can be robust and rate-efficient, their end-to-end latency is often dominated by multi-stage codec processing~\cite{golla2015realtime,wiemann2019rosdraco,alkhalili2022capcd,liu2020mpegcore}, which can be prohibitive for deployment on resource-constrained robotic platforms.

\begin{table*}[t]
\centering
\caption{Representative learned geometry codecs, entropy coding, setup, and runtime.\\
$D$: effective spatial quantization depth ($\equiv$ octree depth / coordinate bit depth).}
\label{tab:merged_entropy_latency}
\footnotesize
\setlength{\tabcolsep}{4pt}
\renewcommand{\arraystretch}{1.10}

\begin{tabular*}{\textwidth}{@{\extracolsep{\fill}}
>{\raggedright\arraybackslash}p{2.35cm}
>{\raggedright\arraybackslash}p{2.35cm}
>{\raggedright\arraybackslash}p{1.95cm}
>{\raggedright\arraybackslash}p{2.15cm}
>{\raggedright\arraybackslash}p{3.10cm}
>{\centering\arraybackslash}p{1.65cm}
>{\centering\arraybackslash}p{1.35cm}@{}}
\toprule
\textbf{Category} 
& \textbf{Entropy Coding} 
& \textbf{Representative} 
& \textbf{Dataset} 
& \textbf{Hardware} 
&\textbf{Total runtime (enc./dec.)}
& \textbf{Entropy encoding} \\
\midrule

Voxel-AR + AC
& \makecell[l]{Context-adaptive\\Arithmetic Coding}
& VoxelDNN~\cite{nguyen2020voxeldnn}
& 8iVFB v2~\cite{deon20178ivfb}
& GeForce RTX 2080
& 2459\,s/6274\,s
& N/R \\

Octree-AR
& Arithmetic Coding
& OctAttention~\cite{fu2022octattention}
& \makecell[l]{SemanticKITTI~\cite{behley2019semantickitti}\\($D{=}14$)}
& NVIDIA V100
& 0.31\,s/321\,s
& N/R \\

Grouped octree
& Arithmetic Coding
& EHEM~\cite{song2023ehem}
& \makecell[l]{SemanticKITTI\\($D{=}14$)}
& NVIDIA V100
& 1.21\,s/1.39\,s
& N/R \\

\makecell[l]{Sparse convolution\\multiscale}
& Arithmetic Coding
& PCGCv2~\cite{wang2021multiscale}
& 8iVFB v2~\cite{deon20178ivfb}
& \makecell[l]{Intel Core i7-8700K\\+ GeForce GTX 1070}
& 1.58\,s/5.40\,s
& N/R \\

\makecell[l]{Sparse multiscale\\occupancy code}
& Arithmetic Coding
& RENO~\cite{you2025reno}
& \makecell[l]{KITTIDetection~\cite{geiger2012we}\\($D{=}14$)}
& \makecell[l]{Xeon Silver 4314\\+ RTX 3090}
& 95\,ms/90\,ms
& 38\,ms \\

\makecell[l]{Octree entropy\\model}
& Range Coding
& OctSqueeze~\cite{huang2020octsqueeze}
& \makecell[l]{N/A ($D{=}14$)}
& \makecell[l]{Intel Xeon E5-2687W\\+ GeForce GTX 1080}
& 91.53\,ms/486.36\,ms
& 3.15\,ms \\

\bottomrule
\end{tabular*}
\end{table*}




\subsection{Learning-Based Point Cloud Compression}
\label{sec:bg_learning}

Learning-based PCC has progressed rapidly by replacing handcrafted probability models with neural networks that capture 3D structure better. Comprehensive surveys categorize modern approaches into voxel/occupancy transforms, octree-structured entropy models, and hybrid pipelines that mix deterministic geometry tools with learned context~\cite{quach2022survey,gao2025_tpami_survey, gao2026volumetricsurvey}. Representative methods learn convolutional transforms for geometry coding~\cite{quach2019learnedtransforms} or variational frameworks for point cloud geometry compression~\cite{wang2019learnedpcgc}. A major line exploits hierarchical occupancy: OctSqueeze introduces octree-structured entropy modeling for LiDAR point clouds~\cite{huang2020octsqueeze}, while VoxelDNN explores learned context models for lossless geometry coding~\cite{nguyen2020voxeldnn}. Despite strong RD performance, many neural codecs remain computation heavy in inference. Moreover, a large portion of the literature focuses on RD curves and reports performance based on powerful GPUs, whereas mobile streaming requires evaluation on constrained devices, including encoder/decoder time and critical-path analysis~\cite{quach2022survey,gao2025_tpami_survey}. This mismatch motivates hybrid designs that place learning only where it yields high impact per compute.

The runtime of a learned geometry codec is also strongly influenced by its entropy model structure: what symbols are coded, how probabilities are generated, and whether decoding can proceed sequentially or in parallel. In other words, two methods may both be ``learning-based PCC'', yet exhibit different decoding cost depending on whether they operate on dense voxels, octree nodes, or sparse multiscale occupancies, and whether their probability model is fully autoregressive or parallelizable. Therefore, to understand the system-level latency implications of prior work, it is necessary to further group representative learned geometry codecs according to their entropy model and decoding dependency structure. We summarize these representative families in Table~\ref{tab:merged_entropy_latency}~\footnote{The runtimes in Table~\ref{tab:merged_entropy_latency} are taken from the respective papers and are not directly comparable because the datasets, quantization settings, and hardware platforms differ. The table is intended to illustrate entropy-model/decoder-dependency trends rather than provide a benchmark comparison.}, together with their reported computing platforms and codec runtimes. Under this view, prior learned geometry codecs can be broadly categorized into four families.

\paragraph{Dense voxel autoregressive occupancy with arithmetic coding (voxel-AR)}
These methods traverse a dense voxel grid and entropy-code voxel occupancies using arithmetic coding, where probabilities are predicted autoregressively. It is intrinsically sequential at the decoder, leading to extreme runtimes. In~\cite{nguyen2021multiscale}, Nguyen \emph{et al.} report VoxelDNN requires 2,459\,s encoding and 6,274\,s decoding on the 8iVFB v2 dataset~\cite{deon20178ivfb}. MSVoxelDNN~\cite{nguyen2021multiscale} reduces this to 54/58\,s by grouping voxels across scales.

\paragraph{Octree node occupancy with large-context neural entropy models (octree-AR)}
Octree-domain approaches avoid dense voxel scanning. Instead, they encode octree occupancy symbols with conditioned context on sibling/ancestor.
However, when the context changes symbol by symbol and the model must be re-evaluated repeatedly during decoding, the decoder becomes the bottleneck.
OctAttention is a representative large-context octree model~\cite{fu2022octattention}.
In the commonly used SemanticKITTI benchmark on a V100 GPU, the re-implementation reported by Song \emph{et al.} shows that OctAttention encodes in 0.31\,s but decodes in 321\,s at depth $D{=}14$. This result reveals a severe encoder--decoder imbalance, where decoding latency is orders of magnitude higher than encoding latency.

\paragraph{Grouped/parallelizable octree entropy coding}
A complementary direction is to redesign the entropy coding so that many symbols can be decoded in parallel, reducing decoder-side latency.
EHEM exemplifies this idea by introducing a grouped context structure~\cite{song2023ehem}.
ECM-OPCC follows the same goal using a multi-group coding strategy to expose parallelism within each layer~\cite{liu2022ecmopcc}.
On SemanticKITTI (V100, $D{=}14$), EHEM reduces decoding to 1.39\,s, while encoding becomes 1.21\,s~\cite{song2023ehem}. These methods demonstrate that ``decoder-fast'' learned octree coding is possible, but the absolute runtime is still in seconds per frame on V100-class GPUs.

\paragraph{Sparse multiscale occupancy with one-shot prediction}
Instead of explicit octree traversal, sparse multiscale methods operate on the active set of occupied voxels across scales and predict occupancy in a more batched/one-shot manner.
SparsePCGC~\cite{wang2021sparsepcgc} and PCGCv2~\cite{wang2021multiscale} are representative sparse-tensor multiscale frameworks. In the re-implementation study of Song \emph{et al.}~\cite{song2023ehem}, SparsePCGC is reported to require 1.76/1.43\,s for encoding/decoding on SemanticKITTI (V100, $D{=}14$).
RENO further pushes this direction to real-time by compressing multiscale sparse occupancy without explicit octree construction
and reporting 0.052/0.050\,s (Enc/Dec) for $D{=}12$ precision KITTI samples on an RTX~3090 and 95/90\,ms for $D{=}14$~\cite{you2025reno}. Recent work has further extended this sparse one-shot family to sequential point cloud compression by incorporating hierarchical inter-frame correlation. HINT~\cite{gao2025hint} introduces temporal cues on top of sparse multiscale occupancy prediction and reports 105/140\,ms encoding/decoding time on 8iVFB v2~\cite{deon20178ivfb} dataset, while achieving up to 43.6\% bitrate reduction relative to G-PCC and outperforming the spatial-only baseline RENO.

Taken together, these families reveal a clear latency hierarchy: voxel-AR and large-context octree-AR methods are dominated by decoder-side sequential dependence, grouped octree methods alleviate but do not eliminate this bottleneck, and sparse multiscale one-shot methods are the most promising direction for real-time deployment. However, even this family still leaves two open issues for mobile settings: runtime on mobile systems can remain prohibitive, and lossless neural entropy coding still requires consistent integer CDF reconstruction across heterogeneous platforms. These issues motivate the design space addressed by \method{}.

\subsection{Entropy Coding in Learned Geometry Codecs}
\label{sec:bg_entropy_backend}

A learned geometry codec typically consists of two coupled components: 
(i) an entropy model that predicts a discrete probability distribution for each occupancy symbol, and 
(ii) an entropy coder that converts these symbol probabilities into a compact bitstream. 
While the entropy model has received substantial attention in prior work, the entropy coding method itself is often treated as a fixed implementation choice. 
In practice, however, this backend directly affects the critical-path latency of real-time geometry compression.

Most prior learned geometry codecs adopt arithmetic coding backends.
For example, the released RENO~\cite{you2025reno} implementation uses \texttt{torchac}~\cite{mentzer2019practical} with normalized integer CDF tables, and earlier learned PCC pipelines also explicitly use arithmetic encoder/decoder in the entropy coding stage.

This design choice is reasonable from a compression perspective.
Arithmetic coding is well suited to symbols with non-uniform probabilities, because it can exploit the predicted distribution with high coding efficiency.
As a result, it has become the default backend in many learned compression systems.
However, its runtime behavior is considerably less favorable for low-latency streaming.
The coding process is inherently sequential at the symbol level, with limited parallelism, so coding latency can be quite high even when the probability model itself is lightweight.

Therefore, for edge-oriented and real-time 3D streaming, an appropriate entropy coding method should be considered.
The key challenge is to preserve compatibility with learned probability outputs while reducing runtime overhead. This motivates us to redesign conventional arithmetic coding into an alternative method based on rANS, which offers a more favorable latency--efficiency trade-off in our deployment-oriented pipeline.

\begin{figure*}[t]
  \centering
  \includegraphics[width=\linewidth]{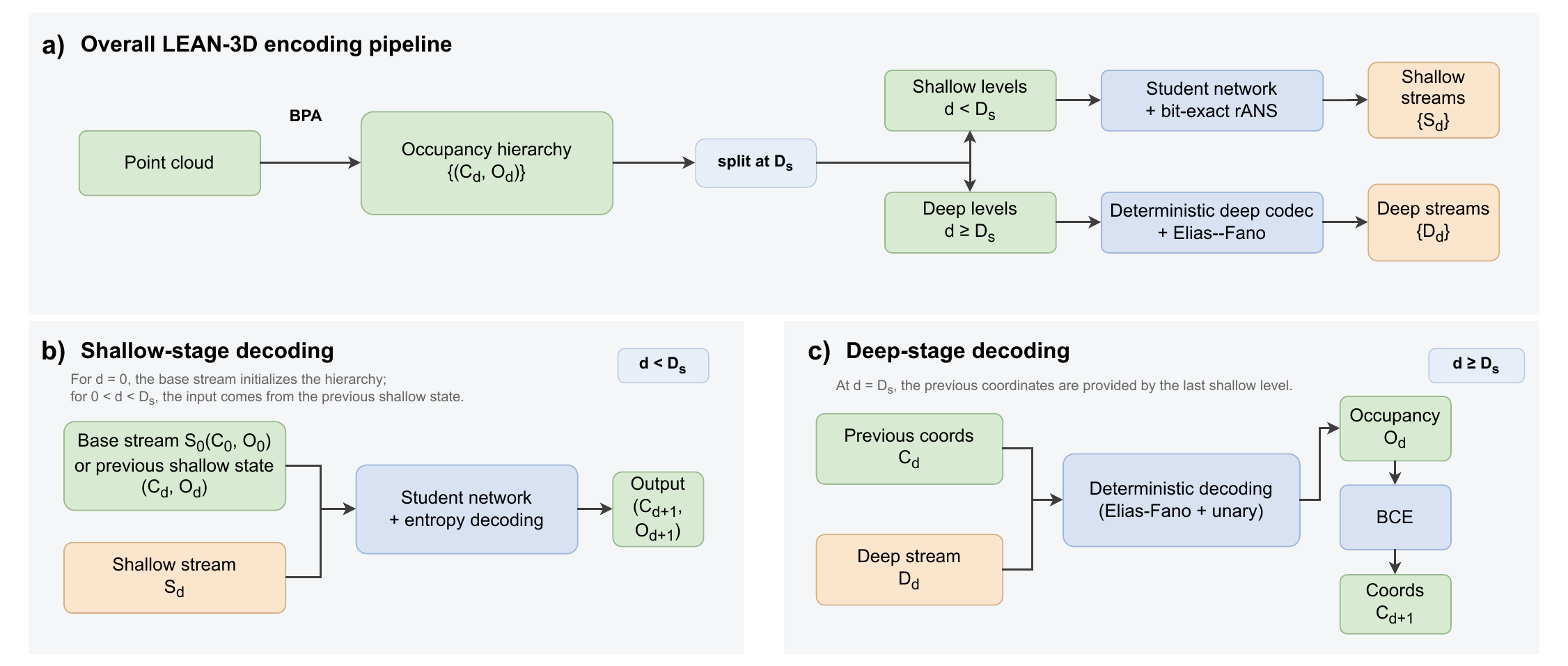}
 \caption{Overview of the proposed \method{} PCC framework. \textbf{(a)} Overall encoding pipeline. The input point cloud is converted into a multiscale occupancy hierarchy by Bitwise Parent Aggregation (BPA) and split at depth $D_s$ into shallow and deep levels. Shallow levels are processed by the student neural network and bit-exact rANS coding to produce shallow streams, while deep levels are handled by Elias--Fano coding to produce deep streams. \textbf{(b)} Shallow-stage decoding. For $d<D_s$, the decoder reconstructs the next shallow level from the base stream or previous shallow level together with the corresponding shallow stream through student neural network and entropy decoding. \textbf{(c)} Deep-stage decoding. For $d\ge D_s$, the decoder combines the previous coordinates with the corresponding deep stream to reconstruct the deep occupancy, and then applies Bitwise Child Expansion (BCE) to generate the next-level active coordinates.}

  \label{fig:system}
\end{figure*}

\section{\method{} PCC Framework}
\label{sec:method}

We present \method{}, a compute-aware hierarchical PCC framework for low-latency point cloud streaming in mobile settings.
This codec is lossless with respect to the quantized coordinates: distortion is only introduced by the input quantization step, while the subsequent encoding/decoding reconstructs the same integer lattice coordinates bit-exactly.
Our design goal is to minimize the critical-path runtime on resource-constrained devices, without sacrificing the probabilistic gains of learned occupancy modeling where it matters most.

\subsection{System Architecture Overview}
\label{sec:method_components}
The overview of \method{} is shown in Fig.~\ref{fig:system}. 
At the system level, \method{} consists of an offline distillation stage and an online codec pipeline. 
As illustrated in Fig.~\ref{fig:system}(a), the online pipeline contains the overall encoding process. The decoder shown in Fig.~\ref{fig:system}(b) and Fig.~\ref{fig:system}(c) is organized into two stages, shallow-stage decoding and deep-stage decoding.

\begin{itemize}
\item \textbf{(Offline) Teacher \& student distillation.}
A pretrained teacher (RENO-style occupancy predictor) provides probability distributions at shallow hierarchy levels.
We distill a lightweight student that preserves performance while reducing computational cost.
\item \textbf{(Online) Edge encoder.}
Given an input frame $\mathcal{P}_t$, the encoder (i) quantizes coordinates, (ii) builds an occupancy pyramid, (iii) encodes shallow occupancies using the student neural network output with rANS coding, and (iv) encodes deep occupancy levels with an Elias--Fano index stream plus compact unary symbols.

\item \textbf{(Online) Decoder (host or edge).}
The decoder reverses the data processing pipeline of the encoder. In our evaluation, we consider both host-side decoding (device-to-host streaming) and edge-side decoding (device-to-device streaming), allowing us to evaluate both latency and cross-platform robustness.
\end{itemize}

\subsection{PCC Encoder and Decoder}
\label{sec:encdec}



 Building on the system overview in Fig.~\ref{fig:system}, we next summarize the frame-level encoder and decoder of \method{}. Fig.~\ref{fig:method_compare}(d) highlights the design position of \method{} as a shallow--deep hybrid codec, while Fig.~\ref{fig:system} shows its end-to-end execution flow. Algorithms~\ref{alg:encoder} and~\ref{alg:decoder} provide a summary of how one input frame is encoded into shallow and deep streams and how these streams are decoded to reconstruct the occupancy hierarchy.

 The encoder first quantizes the input point cloud and constructs a sparse occupancy hierarchy. The hierarchy is then split at depth $D_s$: shallow levels ($d < D_s$) are modeled by the distilled student predictor and encoded into shallow streams using bit-exact rANS coding, whereas deep levels ($d \ge D_s$) are assigned to the deterministic codec through unary/non-unary partitioning and Elias--Fano representation to produce deep streams. 
 
 The decoder follows the same split structure: for $d < D_s$, it reconstructs shallow levels from the base stream and shallow streams using the student neural network and the same bit-exact entropy coding. Once depth $D_s$ is reached, it transitions into deterministic deep-stage decoding, where Elias--Fano coded indices and packed unary symbols are used to recover deep occupancies, followed by Bitwise Child Expansion (BCE) to generate the next-level active coordinates.

The remainder of this section then breaks the pipeline into its main components, including the hierarchy operators Bitwise Parent Aggregation (BPA)/BCE, the shallow learned entropy model, the deterministic deep codec and the bit-exact entropy coding design.

\begin{algorithm}[!t]
\caption{Edge Encoder (per frame)}
\label{alg:encoder}
\begin{algorithmic}[1]
\Require Raw points $\mathcal{P}_t$, quant step $q$, split depth $D_s$
\Ensure Bitstream $\mathbf{b}_t$
\State Quantize $\mathcal{P}_t$ to $\hat{\mathcal{X}}_t$
\State Build occupancy pyramid $\{(C^{(d)},O^{(d)})\}_{d=0}^{L-1}$  
\State Pack base stream: $(C^{(0)},O^{(0)})$
\For{$d=0$ to $D_s-1$} \Comment{Shallow learned}
  \State Compute contexts at $d$ and target features at $d{+}1$
  \State Predict logits for $s_0,s_1$ at $d+1$ and map logits to integer CDFs
  \State rANS-encode ground-truth $(s_0,s_1)$ 
  \State Pack into streams
\EndFor
\For{$d=D_s$ to $L-1$} \Comment{Deep deterministic}
  \State Partition indices into unary $\mathcal{I}_1$ and non-unary $\mathcal{I}_{\ge2}$
  \State Elias--Fano-encode $\mathcal{I}_{\ge2}$
  \State Pack unary child IDs (3 bits each)
  \State Store non-unary occupancy bytes
  \State Append deep chunk stream
\EndFor
\State Pack streams into header + payload and return $\mathbf{b}_t$
\end{algorithmic}
\end{algorithm}

\begin{algorithm}[!t]
\caption{Decoder (host or edge, per frame)}
\label{alg:decoder}
\begin{algorithmic}[1]
\Require Bitstream $\mathbf{b}_t$
\Ensure Reconstructed points $\hat{\mathcal{P}}_t$
\State Parse header and streams
\State Initialize $(C^{(0)},O^{(0)})$ from base stream
\For{$d=0$ to $D_s-1$} \Comment{Shallow learned}
  \State Generate target coordinates $C^{(d+1)}$ from $(C^{(d)},O^{(d)})$
  \State Run student forward to get logits and map logits to integer CDFs 
  \State rANS-decode $(s_0,s_1)$
  \State Reconstruct $O^{(d+1)}=16s_1+s_0$
\EndFor
\For{$d=D_s$ to $L-1$} \Comment{Deep deterministic}
  \State Decode Elias--Fano indices $\mathcal{I}_{\ge2}$
  \State Unpack unary-$k$
  \State Reconstruct $O^{(d)}$
  \State Generate $C^{(d+1)}$ from $(C^{(d)}, O^{(d)})$ by BCE
\EndFor
\State Final BCE expansion yields leaf coordinates $\hat{\mathcal{X}}_t$
\State Inverse quantization: $\hat{\mathcal{P}}_t \leftarrow \hat{\mathcal{X}}_t$ using \texttt{posQ}
\end{algorithmic}
\end{algorithm}

\subsection{Input Quantization and Multiscale Occupancy Codes}
\label{sec:method_repr}
We follow a standard point cloud preprocessing pipeline consisting of quantization and sparse occupancy hierarchy construction. Given a raw point set $\mathcal{P}_t=\{\mathbf{p}_i\in\mathbb{R}^3\}$, we apply uniform quantization with step $q$ (codec parameter \texttt{posQ}):
\begin{equation}
\mathbf{x}_i = \left\lfloor \frac{\mathbf{p}_i}{q} \right\rfloor \in \mathbb{Z}^3,\qquad
\mathcal{X}_t = \{\mathbf{x}_i\}.
\end{equation}
We treat $\mathcal{X}_t$ as the lossless target of the codec.
Geometry is encoded via a dyadic occupancy hierarchy of depth $L$.
Let $d\in\{0,\dots,L-1\}$ index from coarse to fine.
At level $d$, we have a set of active voxel coordinates $C^{(d)} \subset \mathbb{Z}^3$. For each coordinate $\mathbf{v}\in C^{(d)}$, we associate an 8-bit occupancy code: $O^{(d)}(\mathbf{v}) \in \{0,\dots,255\}$.

\subsection{Bitwise Parent Aggregation (BPA)}
\label{sec:method_fog}

A key latency bottleneck in classical pipelines is explicit octree construction.
Instead, we build the hierarchy directly from the set of active voxel coordinates using integer bit operations.

Let $C^{(d+1)}$ denote the active (occupied) coordinates at level $d{+}1$. For each child coordinate $\mathbf{u}=(u_x,u_y,u_z)\in C^{(d+1)}$, we define its parent coordinate as
\begin{equation}
\mathbf{v} = \left\lfloor \frac{\mathbf{u}}{2}\right\rfloor \in \mathbb{Z}^3.
\end{equation}
Let $\boldsymbol{\delta}_k=(\delta_{k,x},\delta_{k,y},\delta_{k,z})$ denote the $k$-th relative child offsets, where $\delta_{k,x},\delta_{k,y},\delta_{k,z}\in\{0,1\}$ for $k=0,\dots,7$.
We define the within-parent child index of $\mathbf{u}$ as
\begin{equation}
k(\mathbf{u}) = (u_x \bmod 2) + 2(u_y \bmod 2) + 4(u_z \bmod 2),
\end{equation}
so that  each child coordinate can be written as:
\begin{equation}
\mathbf{u} = 2\mathbf{v} + \boldsymbol{\delta}_{k(\mathbf{u})}.
\end{equation}
BPA forms the parent set $C^{(d)}$ as the set of unique parent coordinates,
\begin{equation}
C^{(d)} = \big\{\left\lfloor \mathbf{u}/2 \right\rfloor \ \big|\ \mathbf{u}\in C^{(d+1)}\big\},
\end{equation}
and assigns each parent $\mathbf{v}\in C^{(d)}$ an 8-bit occupancy code $O^{(d)}(\mathbf{v})\in\{0,\dots,255\}$, where the $k$-th bit is set to 1
if the $k$-th child position of $\mathbf{v}$ is occupied.
For $k=0,\dots,7$, we define:
\begin{equation}
\begin{gathered}
b_k^{(d)}(\mathbf{v})=
\begin{cases}
1, & \text{if child position } k \text{ of } \mathbf{v} \text{ is occupied},\\
0, & \text{otherwise},
\end{cases}\\[0.4em]
O^{(d)}(\mathbf{v})=\sum_{k=0}^{7} b_k^{(d)}(\mathbf{v})\,2^k.
\end{gathered}
\end{equation}
Thus, bit $k$ of $O^{(d)}(\mathbf{v})$ indicates whether the $k$-th child position of $\mathbf{v}$ is occupied at level $d{+}1$.
 
\begin{figure}[t]
  \centering
  \includegraphics[width=0.8\linewidth,keepaspectratio, trim=.2cm 0.2cm .2cm 0.2cm, clip]{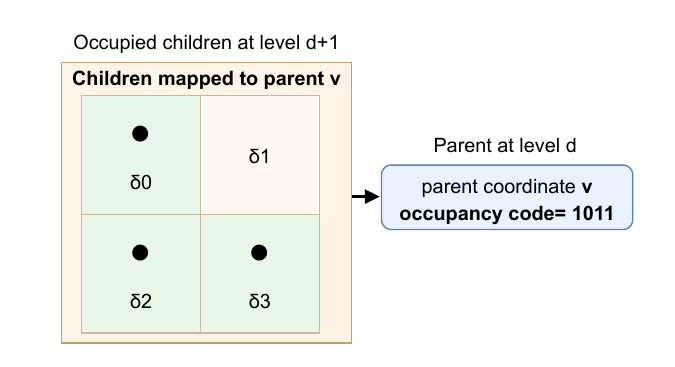}
  \caption{2D toy illustration of BPA: occupied child coordinates at level $d{+}1$ that map to the same parent are grouped into one parent coordinate $\mathbf{v}$ at level $d$, and their within-parent locations are packed into an occupancy code.
}
  \label{fig:bpa}
\end{figure}
In implementation, each child contributes a unique power-of-two bit to its parent. Therefore, per-parent aggregation can be implemented efficiently by grouped summation, yielding an 8-bit occupancy in $[0,255]$.
Fig.~\ref{fig:bpa} illustrates BPA on a simple 2D example: multiple occupied children that share the same parent are aggregated into one parent coordinate together with an occupancy code.
\subsection{Bitwise Child Expansion (BCE)}
\label{sec:method_fcg}
Given level-$d$ active voxel coordinates $C^{(d)}$ and the associated 8-bit occupancy code $O^{(d)}(\mathbf{v})$ for each $\mathbf{v}\in C^{(d)}$, BCE generates the next-level active coordinates by instantiating only the occupied children indicated by set bits:
\begin{equation}
C^{(d+1)}
=
\bigcup_{\mathbf{v}\in C^{(d)}}
\left\{
2\mathbf{v}+\boldsymbol{\delta}_k
\;\middle|\;
k\in\{0,\dots,7\},\;
b_k^{(d)}(\mathbf{v})=1
\right\}.
\label{eq:bce}
\end{equation}

\begin{figure}[t]
  \centering
  \includegraphics[width=0.8\linewidth,keepaspectratio, trim=.2cm 0.2cm .2cm 0.2cm, clip]{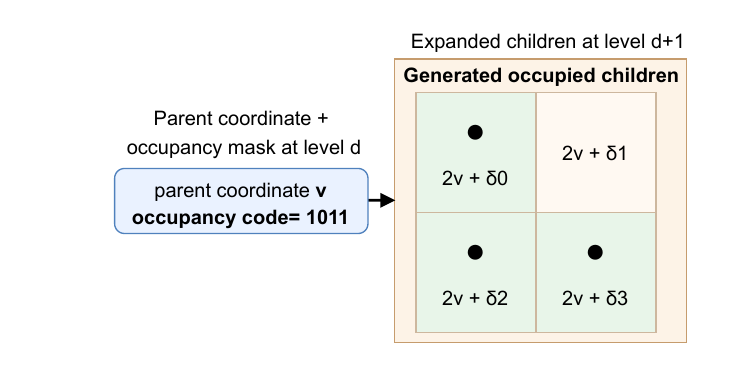}
  \caption{2D toy illustration of BCE: given a parent coordinate $\mathbf{v}$ and its occupancy code, the occupied children are reconstructed with the corresponding offsets at level $d{+}1$.}
  \label{fig:bce}
\end{figure}
In implementation, we enumerate the set bits of the 8-bit mask $O^{(d)}(\mathbf{v})$ and for each set bit $k$ generate one child coordinate as $2\mathbf{v}+\boldsymbol{\delta}_k$, where offset $\boldsymbol{\delta}_k$ is read from a fixed table of eight offsets.
 Fig.~\ref{fig:bce} shows the BCE operation, where the parent coordinate and its occupancy code are expanded back to the occupied children at the next level.
\subsection{Occupancy Statistic Characteristics}
\label{sec:method_split}
 
\begin{figure}[t]
  \centering
  \includegraphics[width=0.8\linewidth]{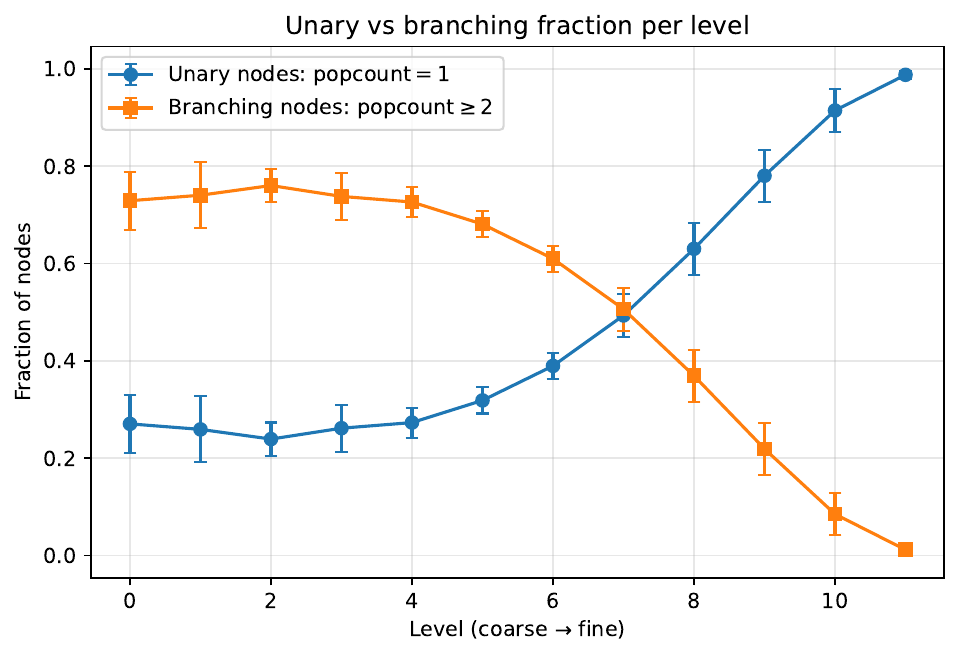}
  \caption{Unary vs.\ branching occupancy statistics across depth levels (coarse to fine), averaged over the KITTIDetection dataset. The unary fraction counts nodes with $\mathrm{popcount}(O)=1$, while the branching fraction counts nodes with $\mathrm{popcount}(O)\geq 2$.}
  \label{fig:unary_branching}
\end{figure}

We observe a shift in occupancy statistics across the hierarchy (Fig.~\ref{fig:unary_branching}). 
At shallow levels, a large fraction of nodes are branching with multiple occupied children, reflecting high structural uncertainty and making learned entropy models most beneficial.
As depth increases, the hierarchy rapidly becomes near-unary: most active nodes expand to a single occupied child:
\begin{equation}
\mathrm{popcount}\!\big(O^{(d)}(\mathbf{v})\big)\approx 1 \quad \text{for large } d.
\label{eq:near_unary_method}
\end{equation}


This pattern implies that the hierarchy should be treated as a compute-allocation problem rather than a fully learned pipeline. Learned modeling is most effective at shallow levels, where it achieves the greatest bit savings per unit of compute, since branching uncertainty is higher and decisions impact a larger portion of the hierarchy. In contrast, deeper levels are increasingly dominated by near-unary expansion and are therefore better handled by a deterministic coding with much lower runtime cost. Accordingly, we introduce a split depth $D_s$: for $d < D_s$ we implement a distilled neural entropy model to capture uncertainty-critical occupancy patterns, whereas for $d \ge D_s$ we adopt a deterministic coding tailored to the near-unary regime.

In practice, $D_s$ is selected based on occupancy statistics.
For each hierarchy level, we compute the fraction of active parent nodes whose occupancy code has a population count of one.
We then define $D_s$ as the first level at which this ratio exceeds 60\%, indicating that the hierarchy has entered a predominantly near-unary regime.
This criterion provides a simple and reproducible rule for identifying the transition from uncertain shallow levels to a predominantly near-unary deep regime.
Once unary-dominant behavior emerges, the gain from learned entropy modeling diminishes, and reallocating deep-level coding from neural prediction to deterministic coding provides a more favorable runtime–rate trade-off.
In all experiments, the resulting $D_s$ is determined offline for each dataset--quantization setting rather than tuned on a per-frame basis.
Investigating adaptive split-selection strategies is left for future work.

\subsection{Learned Occupancy Coding at Shallow Levels}
\label{sec:method_shallow}

\subsubsection{Model design}
\label{sec:method_nibble}
\begin{figure}[t]
  \centering
  \includegraphics[width=\linewidth,keepaspectratio, trim=.2cm 0.5cm .2cm 0.2cm, clip]{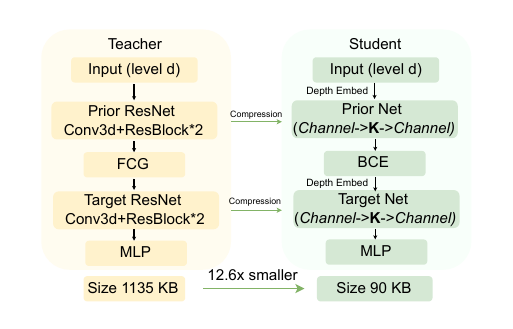}
  \caption{Architecture comparison between the teacher (RENO) model and our distilled student used for shallow levels. The student redesigns the heavy sparse ResNet stacks in both the prior and target branches into a sparse context module (Channel $\rightarrow K \rightarrow$ Channel).}
  \label{fig:nn_comparison}
\end{figure}

As illustrated in Fig.~\ref{fig:nn_comparison}, the teacher adopts two sparse ResNet-style branches (prior and target) with multiple 3D sparse convolutional blocks, which incur substantial neighborhood aggregation cost on edge GPUs due to irregular gather/scatter and memory traffic.
Our student model keeps the same high-level decomposition and cross-scale information flow, but compresses both branches into a single bottleneck sparse context module per level.

Specifically, in each branch we first project node features from $F$ channels to a low-dimensional bottleneck of width $K$ via a $1{\times}1{\times}1$ sparse convolution, applying the only $3{\times}3{\times}3$ neighborhood aggregation at width $K$, and then project back to $F$ channels:
\begin{equation}
\mathbf{f}_{\text{out}}
=
\mathbf{W}_{\uparrow}\,
\phi\!\left(\mathrm{SparseConv}_{3\times3\times3}\!\left(\phi(\mathbf{W}_{\downarrow}\mathbf{f})\right)\right),
\end{equation}
where $\mathbf{f}_{\text{out}}$ is a $F$-channel node feature, and $\mathbf{W}_{\downarrow}$ and $\mathbf{W}_{\uparrow}$ are learnable $1{\times}1{\times}1$ projections for channel reduction ($F\!\rightarrow\!K$, e.g. $F$=32, $K$=8) and expansion ($K\!\rightarrow\!F$).
This design preserves spatial context while reducing the neighborhood aggregation cost from channel width $F$ to $K$. 
The remaining components are kept lightweight and identical in role to the teacher: BCE deterministically generates child coordinates from occupancy codes (Sec.~\ref{sec:method_fcg}), and occupancy code distributions are predicted by small multi-layer perceptrons (MLPs).

We further include depth conditioning to mitigate distribution shift across shallow levels when a single student is shared across multiple depths.
\subsubsection{Distillation objective}
\label{sec:method_distill}
We split each 8-bit occupancy code into two 4-bit sub-symbols, denoted by $s_0$ and $s_1$.
Let $\mathcal{H}_m$ denote the sparse hierarchical context at occupancy code position $m$, and let $\mathbf{p}^{(0)}_m$ and $\mathbf{p}^{(1)}_m$ be the student distributions for $(s_0,s_1)$ at occupancy code position $m$. We use a sequential factorization in both training and inference: $s_0$ is predicted first, and $s_1$ is then predicted conditioned on $s_0$ (i.e., $\mathbf{p}^{(0)}_m=p_\theta^{(0)}(\cdot\mid\mathcal{H}_m)$ and $\mathbf{p}^{(1)}_m=p_\theta^{(1)}(\cdot\mid s_0,\mathcal{H}_m)$), where $\theta$ denotes the parameters of the student model. Let $\mathbf{s}_{m,0}$ and $\mathbf{s}_{m,1}$ denote the corresponding one-hot ground-truth targets. We train the student with a supervised bit-based cross-entropy term on the one-hot ground-truth targets:
\begin{equation}
\mathcal{L}_{\mathrm{CE}}^{\mathrm{bits}}
=
-\sum_m \left(
\mathbf{s}_{m,0}^{\top}\log_2 \mathbf{p}_m^{(0)}
+
\mathbf{s}_{m,1}^{\top}\log_2 \mathbf{p}_m^{(1)}
\right).
\end{equation}
This term ensures that the student predicts the correct ground-truth occupancy symbols.
To further transfer the teacher's soft distributional knowledge to the simplified student, we add a Kullback--Leibler (KL) divergence distillation term, converted to bits by dividing by $\log 2$, to match the teacher distributions $\mathbf{q}^{(h)}_m$:
\begin{equation}
\mathcal{L}_{\mathrm{KD}}^{\mathrm{bits}}
=\sum_{h\in\{0,1\}}\sum_{m} \mathrm{KL}\!\left(\mathbf{q}^{(h)}_m \,\|\, \mathbf{p}^{(h)}_m\right)\big/\log 2.
\end{equation}
Unlike the cross-entropy term, which uses hard ground-truth labels, this term encourages the student to match the teacher's soft output distributions.
Here, $h \in \{0,1\}$ indexes the two 4-bit occupancy groups corresponding to $s_0$ and $s_1$, respectively. The final objective is $\mathcal{L}=\mathcal{L}_{\mathrm{CE}}^{\mathrm{bits}}+\lambda\,\mathcal{L}_{\mathrm{KD}}^{\mathrm{bits}}$.
\subsection{Deterministic Coding at Deep Levels}
\label{sec:method_deep}

For deep levels $d \ge D_s$, we avoid neural inference and exploit the near-unary structure.
Let $\{\mathbf{v}_i\}_{i=1}^{N_u} \subset C^{(d)}$ denote the active voxel coordinates at the current deep level in coding order, where
$N_u = |C^{(d)}|$ is the number of active voxels at that level, and let
$O_i^{(d)} = O^{(d)}(\mathbf{v}_i)$ denote the corresponding occupancy code.
For the current deep level $d$, we partition the index set into unary and non-unary subsets
$\mathcal{I}_1$ and $\mathcal{I}_{\ge 2}$, corresponding to voxels with a single child and voxels with multiple children, respectively:
\begin{equation}
\begin{aligned}
\mathcal{I}_1 &= \{i : \mathrm{popcount}(O_i^{(d)}) = 1\},\\
\mathcal{I}_{\ge 2} &= \{i : \mathrm{popcount}(O_i^{(d)}) \ge 2\}.
\end{aligned}
\end{equation}

\textbf{Non-unary indices via Elias--Fano:} We store the sorted index set $\mathcal{I}_{\ge 2} \subset \{1,\dots,N_u\}$ using Elias--Fano encoding, which compactly represents monotone sequences and supports fast decoding.
For each $i\in\mathcal{I}_{\ge2}$ we additionally store the full 8-bit occupancy value $O_i^{(d)}$ following the sorted order of $\mathcal{I}_{\ge 2}$.

 \textbf{Unary nodes as 3-bit child IDs:} For $i\in\mathcal{I}_{1}$ we have $O_i^{(d)} = 2^{k_i}$ with $k_i\in\{0,\dots,7\}$.
We store $\{k_i\}$ as a packed fixed-length sequence using 3 bits per symbol.

This codec removes neural inference from the deep levels of the hierarchy and makes the decoding of deep levels lightweight and robust.

\subsection{Bit-Exact Entropy and rANS Coding}
\label{sec:method_bitexact}
\textbf{Consistent CDF construction:} Given a logit vector $\mathbf{z}\in\mathbb{R}^{16}$, we first quantize it into integers:
\begin{equation}
\tilde{\mathbf{z}}=\left\lfloor 128\,\mathbf{z}+0.5 \right\rfloor \in \mathbb{Z}^{16}.
\end{equation}

We then define a consistent ranking score for each symbol:
\begin{equation}
s_r = \lambda\,\tilde{z}_r - r,\qquad r=0,\dots,15,
\end{equation}
where $\tilde{z}_r$ denotes the $r$-th entry of $\tilde{\mathbf{z}}$ and $\lambda$ is a fixed positive constant. In our implementation, we set $\lambda=1000$, which ensures that the ordering is dominated by the quantized logit value $\tilde{z}_r$, while the symbol index $r$ is used only to break ties. Sorting $\{s_r\}_{r=0}^{15}$ in descending order gives a permutation $\pi$, where $\pi(j)$ denotes the symbol ranked at position $j$.
Next, we introduce a fixed count vector:
\begin{equation}
\mathbf{n}^{\star}=(60000,\underbrace{369,\dots,369}_{14\ \text{times}},370)\in\mathbb{N}^{16},
\end{equation}
where $\mathbf{n}^{\star}=(n_0^{\star},\dots,n_{15}^{\star})$ denotes the template count vector ordered by rank, and $n_j^{\star}$ is its $j$-th entry. Its total mass satisfies:
\begin{equation}
\sum_{j=0}^{15} n_j^\star = 65536 = 2^{16},
\end{equation}
as required by rANS.

Let $\mathbf{n}=(n_0,\dots,n_{15})\in\mathbb{N}^{16}$ denote the final symbol-count vector in the original symbol order, where $n_r$ is the count assigned to symbol $r$. The template is assigned according to the rank order:
\begin{equation}
n_{\pi(j)} = n_j^{\star},\qquad j=0,\dots,15.
\end{equation}
Therefore, the highest-ranked symbol receives count $60000$, the lowest-ranked symbol receives count $370$, and all remaining symbols receive count $369$. This template is designed to preserve a dominant count for the highest-ranked symbol while assigning nonzero mass to all symbols. Its purpose is cross-platform consistency and low-latency coding rather than exact approximation of the original softmax probabilities.

Finally, the integer CDF is obtained by prefix summation over $\mathbf{n}$:
\begin{equation}
\mathrm{CDF}_0 = 0,\qquad
\mathrm{CDF}_{r+1} = \sum_{t=0}^{r} n_t,\qquad r=0,\dots,15.
\end{equation}
All steps after logit quantization are integer-only operations, which guarantee bit-exact CDF construction across different devices.

\textbf{Entropy encoder:} We encode each shallow level using rANS with the generated CDFs for the two 4-bit sub-symbols $s_0$ and $s_1$. Compared to arithmetic coding, rANS supports lower-latency implementations, which fit well with our objective.

\section{Implementation and Evaluation}
\label{sec:exp}
We conducted comprehensive experiments to evaluate \method{} through both the developed working prototype and simulations for mobile point cloud streaming, with emphasis on practical deployment metrics rather than bitrate alone. Our goal is not to claim bitrate minimization in all settings, but to evaluate whether \method{} achieves a more favorable runtime--rate--robustness operating point for mobile 3D streaming. The experiments are designed from four perspectives: 
(i) full codec runtime and component-wise latency attribution of the proposed dual-codec, 
(ii) compression and transmission behavior under different bandwidth conditions, and their impact on end-to-end streaming performance,
(iii) generalization across datasets and heterogeneous edge-to-host decoding, and (iv) edge-side energy efficiency.

\begin{figure}[htbp]
  \centering
  \includegraphics[width=0.8\linewidth,keepaspectratio, trim=.4cm 0.3cm 0.3cm 0.32cm, clip]{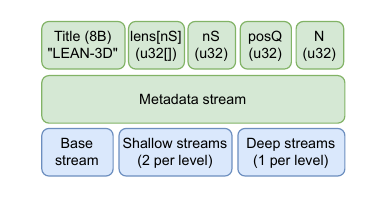}
  \caption{Packet structure of one encoded frame. The top row shows the fixed header, including the file signature, stream-length array, stream count, and frame-level fields. The payload then consists of a metadata stream, a base stream, shallow streams, and deep streams concatenated in a fixed order.}
  \label{fig:packet}
\end{figure}

\textbf{Packet format:} For deployment, each encoded frame is serialized into the packet format shown in Fig.~\ref{fig:packet}. 
The packet consists of a fixed header followed by payload streams concatenated in a fixed order so that the decoder can recover each component.
We organize the streams as follows:
\begin{enumerate}
    \item \textbf{Header}: stores the file signature \texttt{Title} (8B), the stream-length array \texttt{lens[nS]} (\texttt{u32[]}), the stream count \texttt{nS} (\texttt{u32}), and the frame-level fields \texttt{posQ} and \texttt{N} (\texttt{u32}).

    \item \textbf{Metadata stream}: stores the hierarchy-depth settings \texttt{depths} and \texttt{shallow\_D}, together with the integer entropy-coding parameters \texttt{fp\_inv\_step}, \texttt{fp\_B}, and \texttt{fp\_KMAX}.
    
    \item \textbf{Base stream}: stores the coarsest-scale representation after downscaling, including \texttt{n0} (u32), \texttt{base\_xyz} (int32$\times 3 \times n0$), and \texttt{base\_occ} (u8$\times n0$).
    
    \item \textbf{Shallow-level streams}: for each shallow hierarchy level $d < D_s$, two rANS streams are transmitted, corresponding to the binary occupancy groups $s_0$ and $s_1$.
    
    \item \textbf{Deep-level streams}: for each deeper level $d \ge D_s$, one deep stream is transmitted. 
    Each deep stream contains the associated metadata 
    (\texttt{Nu}, \texttt{Msplit}, \texttt{L}, and the byte lengths of the encoded substreams), 
    followed by the corresponding payload bytes for Elias--Fano high/low parts, packed unary-$k$, and non-unary occupancy symbols.
\end{enumerate}

\textbf{Datasets:} Unless otherwise stated, the main experiments are conducted on KITTIDetection over 300 consecutive frames from each selected sequence.
We further use Argoverse2 (AV2)~\cite{wilson2023argoverse2} and SemanticKITTI (SK) for cross-dataset validation in order to test whether the observed runtime--rate performance generalizes beyond a single LiDAR point cloud distribution.
\begin{figure}[htbp]
  \centering
  \includegraphics[width=\linewidth]{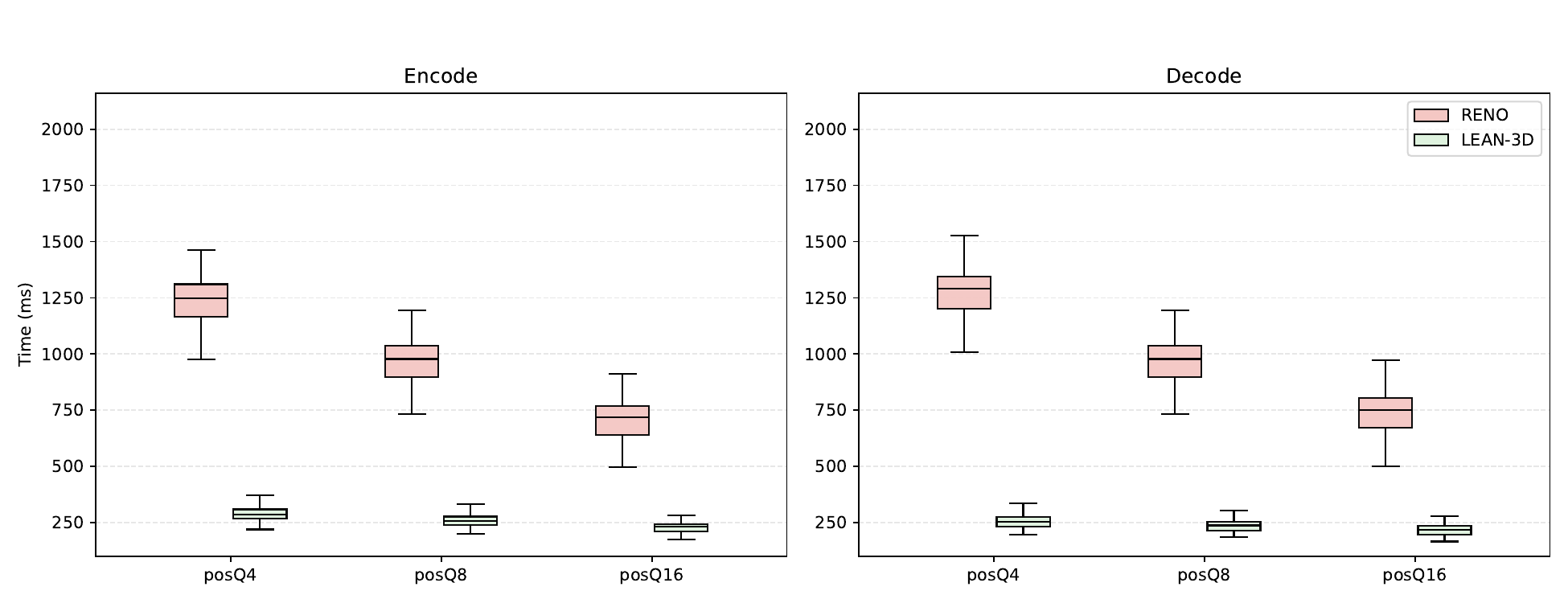}
  \caption{Per-frame encoding and decoding time comparison of RENO and \method{} on Jetson Orin Nano under different quantization settings ($\texttt{posQ}\in\{4,8,16\}$) on KITTIDetection.}
  \label{fig:enc_dec_compare}
\end{figure}

\begin{figure*}[!t]
  \centering
  \includegraphics[width=\linewidth]{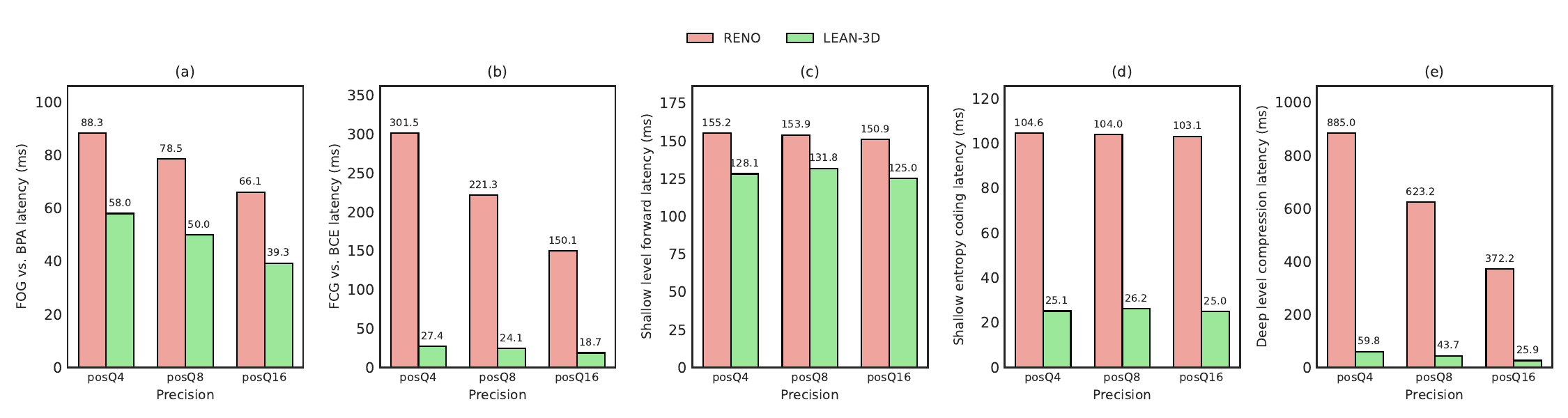}
  \caption{Component-wise latency attribution of RENO and \method{} on Jetson Orin Nano under different quantization settings ($\texttt{posQ}\in \{4, 8, 16\}$) on KITTIDetection. \textbf{(a)} Fast Occupancy Generator (FOG) vs. BPA latency. \textbf{(b)} Fast Coordinates Generator (FCG) vs. BCE latency. \textbf{(c)} Shallow level forward latency. \textbf{(d)} Shallow entropy encoding latency. \textbf{(e)} Deep level compression latency.}
  \label{fig:time_breakdown}
\end{figure*}
\textbf{Prototype setup:} The edge-side experiments are conducted on an NVIDIA Jetson Orin Nano 8GB Developer Kit, representing a resource-constrained mobile platform for point cloud geometry compression.
The host-side experiments are conducted on a desktop equipped with an Intel Core i9-14900 CPU and an NVIDIA RTX 4090 GPU.
Unless otherwise specified, edge-side latency results are measured on the Jetson platform, while heterogeneous deployment experiments use the Jetson as encoder and the desktop host as decoder.

\textbf{Split-depth setting:} Using the criterion in Sec.~\ref{sec:method_split}, the selected split depths are fixed for each dataset--quantization setting.
For KITTIDetection and SemanticKITTI, we use $D_s=\{4,3,2\}$ for $\texttt{posQ}\in\{4,8,16\}$, respectively.
For Argoverse2, we use $D_s=\{6,5,4\}$ for $\texttt{posQ}\in\{4,8,16\}$, respectively.
This trend is expected because a larger $\texttt{posQ}$ yields a coarser voxelization and thus a shallower occupancy hierarchy, so the selected split depth $D_s$ decreases accordingly.

\subsection{Full Encode/Decode Runtime}

We first evaluate full-frame encode and decode runtime, because continuous streaming on mobile systems is fundamentally bounded by the service time of the codec at both ends. Fig.~\ref{fig:enc_dec_compare} reports per-frame runtime distributions under different quantization settings on KITTIDetection.

Fig.~\ref{fig:enc_dec_compare} shows that the benefit of \method{} extends beyond lower average runtime. Across the tested quantization settings, \method{} achieves approximately $3.2\times$--$4.3\times$ faster encoding and $3.5\times$--$5.1\times$ faster decoding than RENO, while also exhibiting a tighter runtime distribution across frames. This indicates not only lower per-frame latency but also lower runtime variability, which is important for streaming systems because queue buildup and deadline misses depend on jitter as well as mean processing time. The results therefore suggest that \method{} provides both a smaller per-frame processing budget and a more stable runtime profile for long-sequence mobile deployment.


\subsection{Component-Wise Latency Attribution}

To understand where the runtime gain comes from, we next break the encoder side into the major stages on the practical critical path: occupancy generation, context construction, shallow-level inference, shallow entropy coding, and deep-level compression. The purpose of this analysis is to evaluate whether \method{} improves the overall pipeline or merely accelerates one isolated module.

Fig.~\ref{fig:time_breakdown} shows that the gain is distributed across the pipeline but is concentrated in the stages that dominate the execution path of the baseline. 
The main reduction does not come only from shrinking the shallow predictor. 
At $\texttt{posQ}=4$, the BCE stage is accelerated by 11$\times$ (from 301.5\,ms to 27.4\,ms in Fig.~\ref{fig:time_breakdown}(b)), shallow entropy coding is accelerated more than 4$\times$  (from 104.6\,ms to 25.1\,ms in Fig.~\ref{fig:time_breakdown}(d)), and deep-level compression achieves a speedup of 14.8$\times$ (from 885.0\,ms to 59.8\,ms in Fig.~\ref{fig:time_breakdown}(e)), whereas shallow forward latency has modest acceleration from 155.2\,ms to 128.1\,ms (Fig.~\ref{fig:time_breakdown}(c)). Similar patterns persist at $\texttt{posQ}=8$ and $\texttt{posQ}=16$. In short, the largest runtime acceleration comes from BCE, shallow entropy coding, and deep-level compression, which are the dominant runtime bottlenecks of the baseline. The speedup is primarily driven by restructuring the heavy deep-hierarchy path as a deterministic stage.

\begin{figure*}[!t]
  \centering
  \includegraphics[width=\linewidth]{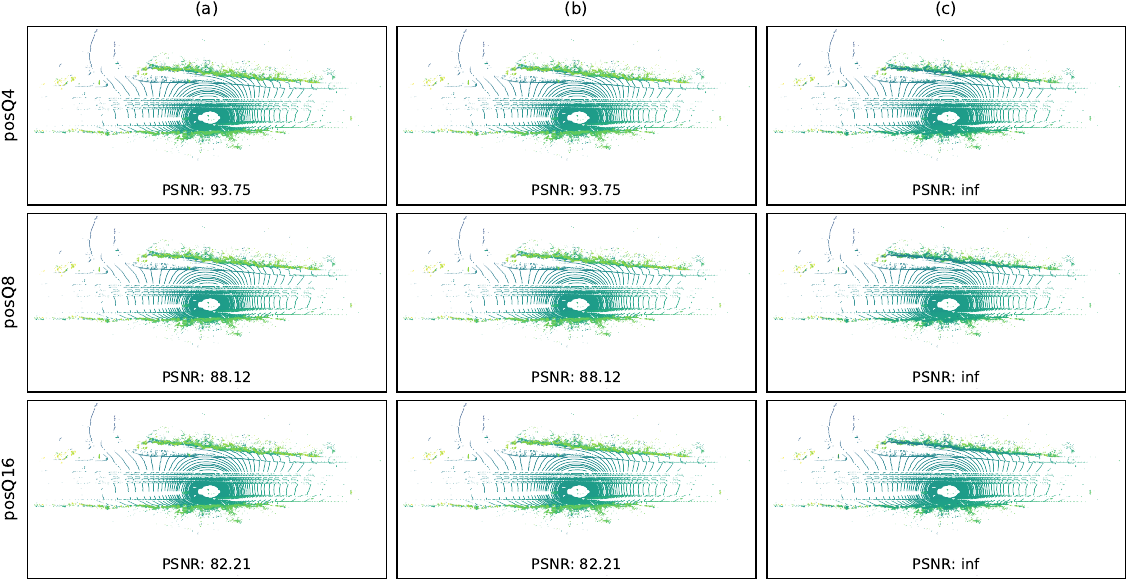}
  \caption{Qualitative reconstruction comparison under different quantization settings ($\texttt{posQ}\in\{4,8,16\}$) on KITTIDetection, where (a), (b), and (c) correspond to RENO, \method{}, and ground truth, respectively. The reported PSNR is computed against the original raw point cloud. The identical PSNR values of RENO and \method{} indicate that both codecs reconstruct the same quantized geometry, while the distortion is entirely introduced by the input quantization.}
  \label{fig:compare33}
\end{figure*}
\subsection{Qualitative Reconstruction Analysis}

We next verify that the runtime acceleration does not come at the cost of visible structural degradation beyond the chosen quantization level. Fig.~\ref{fig:compare33} compares qualitative reconstructions of \method{} and RENO under the tested settings on KITTIDetection.

The two codecs preserve the same quantized structure at each setting, while the visible loss is caused by the quantization step itself. This is consistent with the goal of \method{}: to improve the runtime--rate trade-off of the codec while preserving the target quantized geometry. 

\subsection{Bandwidth-Limited End-to-End Evaluation}
\label{sec:exp_e2e_bandwidth}

\begin{figure}[htbp]
  \centering
  \includegraphics[width=\linewidth]{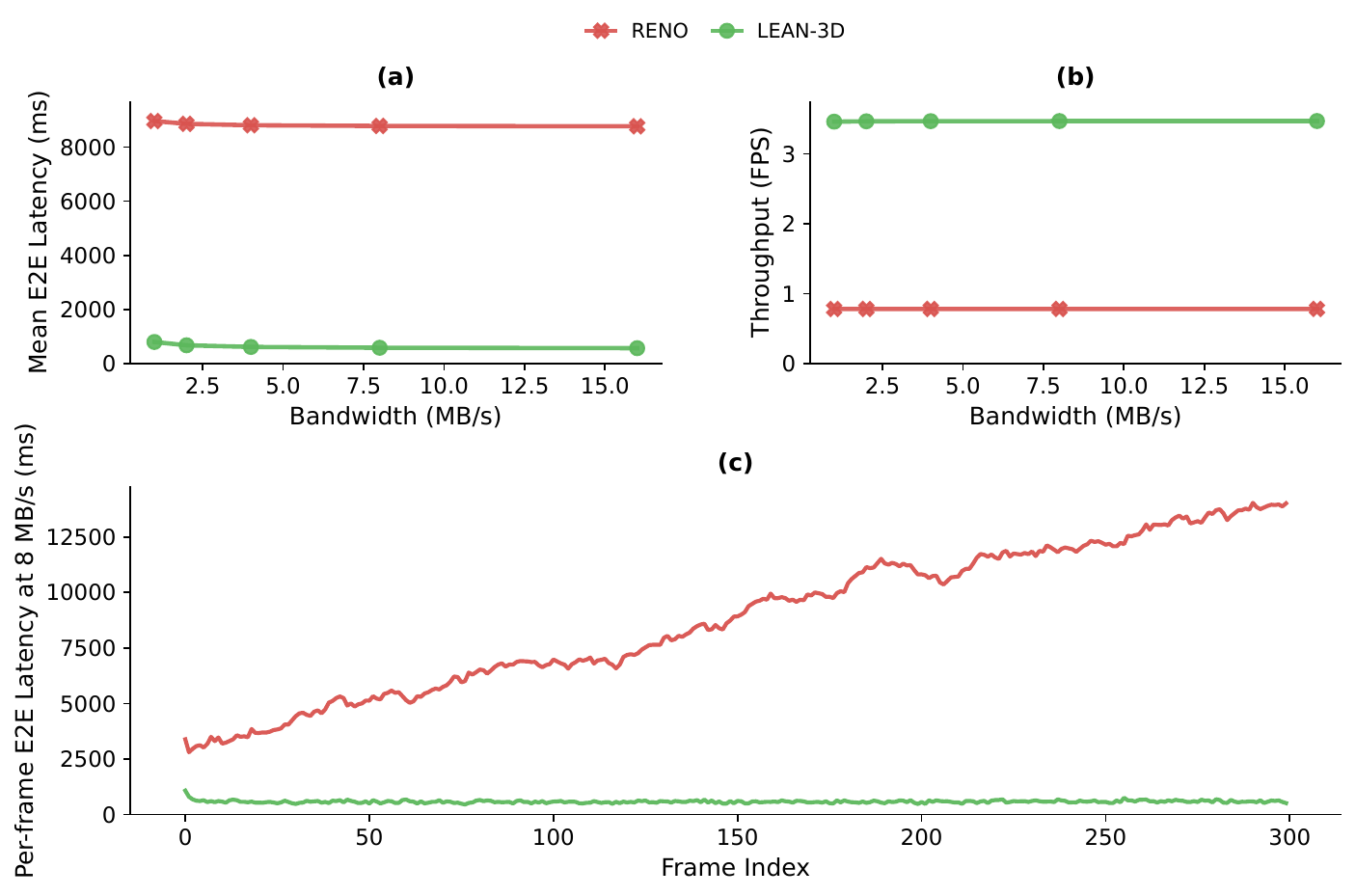}
  \caption{System-level streaming performance under $\texttt{posQ}=4$ on KITTIDetection. \textbf{(a)} Mean end-to-end latency versus available bandwidth. \textbf{(b)} Output throughput versus available bandwidth. \textbf{(c)} Per-frame end-to-end latency at 8\,MB/s, illustrating the difference in queue buildup over time.}
  \label{fig:e2e_perf}
\end{figure}

A faster codec is only useful for streaming if the runtime gain persists once computation and communication are coupled. We therefore perform a trace-driven streaming simulation under bandwidth constraints, using the heaviest-load setting $\texttt{posQ}=4$. Fig.~\ref{fig:e2e_perf} reports the resulting latency and throughput performance.

For each codec, we first record three frame-level traces over 300 frames: measured encoding time $t_i^{\mathrm{enc}}$, compressed payload size $b_i$, and measured decoding time $t_i^{\mathrm{dec}}$. These traces are then replayed through a sequential three-stage pipeline,
\[
\text{encoder} \rightarrow \text{bandwidth-limited link} \rightarrow \text{decoder},
\]
which matches the execution policy of the current prototype. The pipeline is modeled as First-Come, First-Served (FCFS) with network service time: $t_i^{\mathrm{net}}=\frac{b_i}{B}$, where $B$ is the effective bandwidth. We report the per-frame streaming latency from the start of encoding to the completion of decoding. This setup intentionally isolates the interaction between measured codec cost and bandwidth limitation, without introducing additional protocol-layer effects that would affect both codecs.

Fig.~\ref{fig:e2e_perf}(a) shows that \method{} achieves lower mean end-to-end latency than RENO across the tested bandwidth range. Fig.~\ref{fig:e2e_perf}(b) shows that \method{} achieves a higher decoding throughput than RENO, indicating that the pipeline can deliver completed frames more efficiently. 
More importantly, \method{} keeps the combined compute--communication pipeline farther from the unstable operating region. This is most evident in Fig.~\ref{fig:e2e_perf}(c): under 8\,MB/s, the latency of RENO increases progressively over time, indicating queue buildup, whereas the latency of \method{} remains nearly stable. Therefore, the advantage of the proposed design is not only lower average latency, but also a reduced risk of entering a backlog regime during continuous streaming.

\subsection{Rate Trade-Off of the Hierarchical Design}

\begin{figure}[htbp]
  \centering
\includegraphics[width=\linewidth]{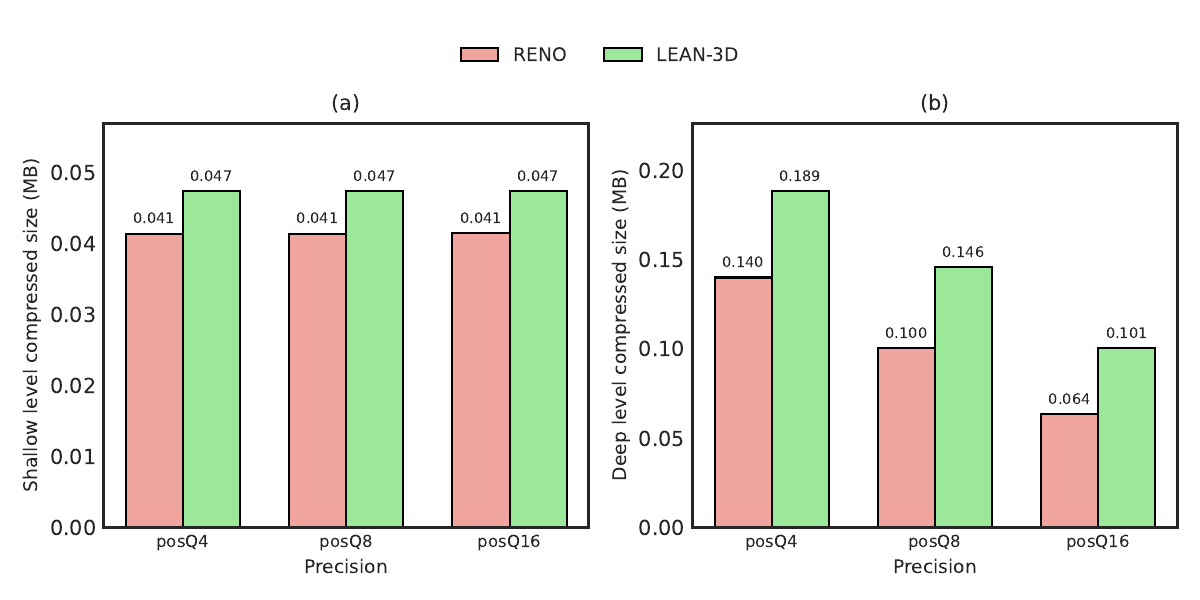}
  \caption{Compressed size breakdown of RENO and \method{} under different quantization settings ($\texttt{posQ}\in\{4,8,16\}$) on KITTIDetection. \textbf{(a)} Shallow level compressed size. \textbf{(b)} Deep level compressed size.}
  \label{fig:size_breakdown}
\end{figure}

The runtime advantage of \method{} introduces a measurable rate overhead, making it important to explicitly characterize the associated trade-offs. We therefore separate the compressed size of the shallow and deep branches and analyze them independently in Fig.~\ref{fig:size_breakdown}.

The results show that \method{} operates at a different compute--rate point from the baseline. The shallow-stream overhead is small and nearly constant across quantization settings: The shallow part changes only from 0.041\,MB to 0.047\,MB. The main increase comes from the deterministic deep branch, where the compressed size changes from 0.140\,MB to 0.189\,MB at $\texttt{posQ}=4$, from 0.100\,MB to 0.146\,MB at $\texttt{posQ}=8$, and from 0.064\,MB to 0.101\,MB at $\texttt{posQ}=16$. 

\method{} introduces a bounded overhead concentrated in the deep regime in exchange for a significantly simpler execution path. For mobile systems, this trade-off is often preferable, as a modest increase in data size may be acceptable if it reduces codec delay and persistent queue buildup during continuous streaming. 

\subsection{Cross-Dataset Evaluation}
\label{sec:exp_cross_dataset}

\begin{figure}[htbp]
  \centering
  \includegraphics[width=\linewidth]{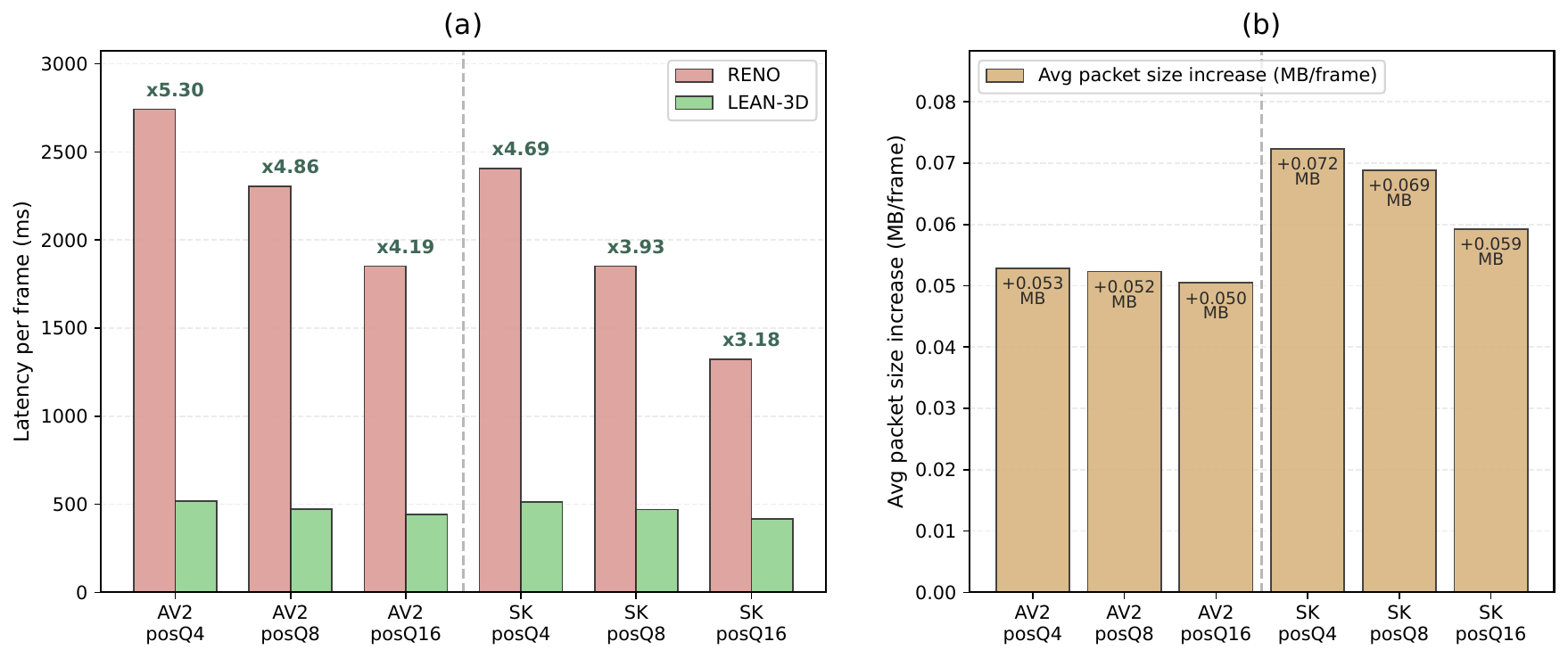}
  \caption{Cross-dataset validation on Argoverse2 (AV2) and SemanticKITTI (SK). \textbf{(a)} Edge-side mean encoding latency under different quantization settings ($\texttt{posQ}\in\{4,8,16\}$). \textbf{(b)} Corresponding average packet-size increase of \method{} relative to RENO.}
  \label{fig:cross_dataset}
\end{figure}

A viable mobile codec should generalize across diverse scene distributions, rather than depending on a particular one. We therefore repeat the same evaluation beyond KITTIDetection on two additional LiDAR benchmarks, Argoverse2 (AV2) and SemanticKITTI (SK), and summarize the results in Fig.~\ref{fig:cross_dataset}.

Fig.~\ref{fig:cross_dataset}(a) shows that the latency advantage of \method{} is consistent across datasets and quantization levels. On AV2, \method{} reduces per-frame latency by $5.30\times$, $4.86\times$, and $4.19\times$ under $\texttt{posQ}=4,8,16$, respectively. On SemanticKITTI, the speedup remains at $4.69\times$, $3.93\times$, and $3.18\times$. Fig.~\ref{fig:cross_dataset}(b) shows that this acceleration is achieved with only a modest compressed data size increase: approximately 0.05--0.053\,MB per frame on AV2 and 0.059--0.072\,MB per frame on SemanticKITTI. 
These results indicate that the benefit of \method{} is not tied to one specific dataset or point-density pattern. Instead, the gain is driven by the codec architecture itself: learned computation is applied where occupancy uncertainty is high, whereas the latency-heavy deep regime is governed by deterministic coding.

\subsection{Heterogeneous Platform Evaluation}
\label{sec:heterogenous}

Cross-platform robustness is another requirement for practical deployment. In real systems, the encoder and decoder often run on heterogeneous hardware, for example an embedded edge device transmitting to a desktop host. In such settings, lossless learned entropy coding requires the encoder and decoder to reconstruct exactly the same integer CDFs. Otherwise, arithmetic decoding can fail even if the model architecture is identical. An accompanying video is available online to illustrate the cross-platform edge-to-host prototype and its comparison with RENO~\footnote{\href{https://youtu.be/ht_J4KiG2z0}{Demonstration video}: \url{https://youtu.be/ht_J4KiG2z0}}.

\begin{figure}[htbp]
  \centering
  \includegraphics[width=\linewidth]{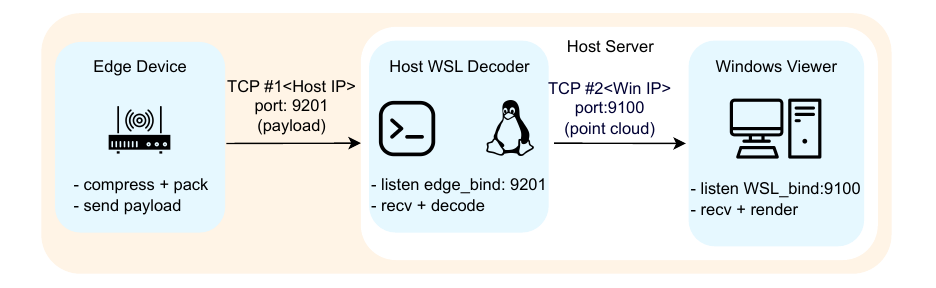}
  \caption{Edge-to-host prototype setup. The edge device compresses and sends point cloud packets to a Windows Subsystem for Linux (WSL)-based host decoder over TCP, and the decoded point cloud is then forwarded to the Windows viewer for rendering.}
  \label{fig:edge_host_setup}
\end{figure}

We first test direct heterogeneous deployment of RENO using the setup in Fig.~\ref{fig:edge_host_setup} under a 40\,MB/s link. 
\begin{figure}[htbp]
  \centering
  \includegraphics[width=\linewidth]{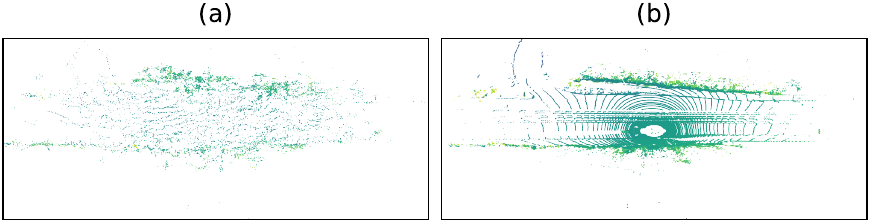}
  \caption{Decoding failure of cross-platform deployment of RENO on KITTIDetection. \textbf{(a)} Invalid reconstruction at host when the bitstream is encoded on edge device using original RENO pipeline. \textbf{(b)} Ground truth.}
  \label{fig:failure}
\end{figure}
Without the bit-exact entropy design in Sec.~\ref{sec:method_bitexact}, direct heterogeneous deployment of RENO exhibits a 100\% decoding failure rate in this setup. Fig.~\ref{fig:failure} shows a representative failure case: the bitstream is produced by a nominally valid encoding pipeline, but the host-side reconstruction becomes invalid because encoder and decoder do not rebuild the same entropy tables. This experiment highlights that entropy consistency is crucial in heterogeneous mobile deployment.

\begin{figure}[htbp]
  \centering
  \includegraphics[width=\linewidth]{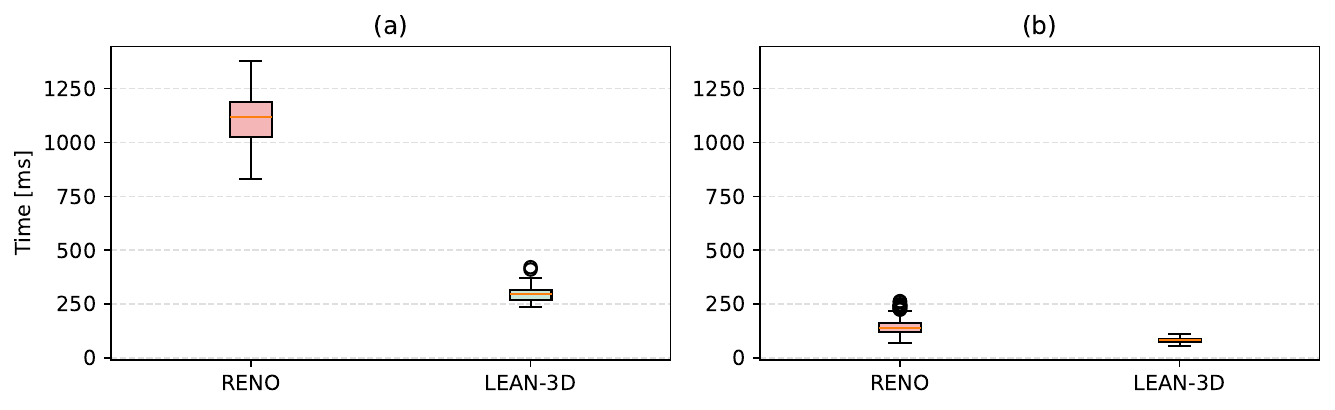}
  \caption{Edge--host codec latency on KITTIDetection under heterogeneous execution (Jetson Orin Nano as encoder, desktop host as decoder). \textbf{(a)} Encoding latency comparison on the edge. \textbf{(b)} Decoding latency comparison on the host.}
  \label{fig:hetero_lat}
\end{figure}

After integrating the proposed bit-exact entropy coding, the encoder and decoder generate consistent integer CDFs across platforms, enabling correct lossless decoding on the tested sequences. We then benchmark edge-to-host codec latency in Fig.~\ref{fig:hetero_lat}.

The heterogeneous results show that \method{} accelerates encoding on the resource-constrained edge while also delivering lower and more stable decoding latency on the host. Together with the failure analysis above, this confirms that both the proposed bit-exact entropy coding and compute-aware split are necessary for a deployable cross-platform prototype setup.

\subsection{Energy Consumption}
\label{sec:energy}

\begin{figure}[htbp]
  \centering
  \includegraphics[width=\linewidth]{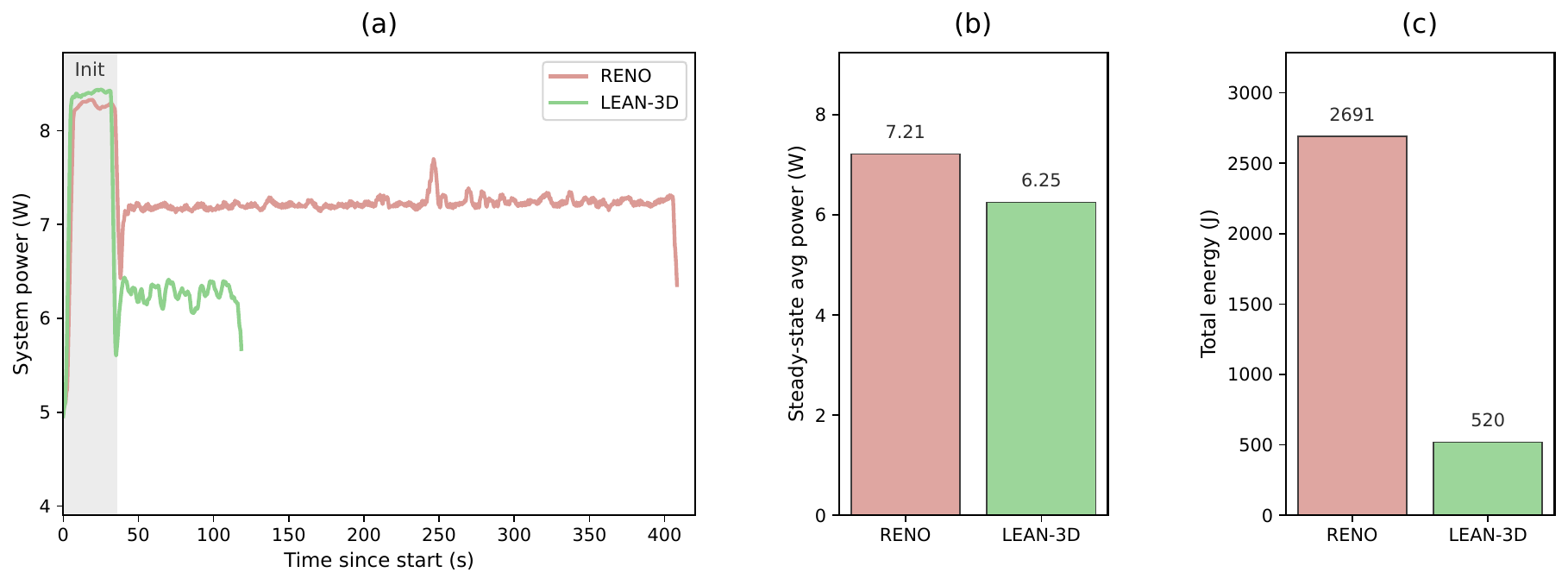}
  \caption{Energy comparison on Jetson Orin Nano for KITTIDetection at $\texttt{posQ}=4$. \textbf{(a)} System power trace (VDD\_IN) during continuous compression. The gray-shaded region denotes the one-time system initialization stage. \textbf{(b)} Steady-state average power. \textbf{(c)} Energy consumption (without initialization) over the full run.}
  \label{fig:energy}
\end{figure}

Beyond runtime, energy is another constraint for portable devices. We therefore measure the edge-side power trace during continuous compression on KITTIDetection at $\texttt{posQ}=4$ and report both steady-state average power and total energy in Fig.~\ref{fig:energy}.

As shown in Fig.~\ref{fig:energy}(a) and Fig.~\ref{fig:energy}(b), \method{} reduces the steady-state average power from 7.21\,W for RENO to 6.25\,W. More importantly, because the codec completes the same workload much faster, the cumulative energy consumption drops from 2691\,J to 520\,J, corresponding to a $5.1\times$ reduction, as shown in Fig.~\ref{fig:energy}(c).

This distinction is important for mobile deployment. While a lower system power is desirable, for continuous point cloud streaming it is a crucial requirement to reduce the total energy consumption. By shortening the codec critical path on the edge device, \method{} improves both responsiveness and energy efficiency, making 3D streaming feasible on energy-constrained platforms.
 
\section{Discussion and Future Work}
\label{sec:discussion}

This work is motivated by a deployment reality that is often under emphasized in the PCC literature: for mobile 3D streaming, an optimal operating point is not necessarily the one with the lowest bitrate, but the one that delivers the most favorable runtime--rate--robustness trade-off under constrained hardware. Our results show that a latency-oriented redesign can improve the practical operating region of a learned geometry codec. In particular, the proposed shallow--deep dual codec reduces the critical-path computation at both encoding and decoding, improves runtime predictability, and mitigates queue buildup under bandwidth-limited streaming. 

A second implication is that the hierarchy itself should be viewed as a compute allocation structure, not only a representation structure. Our experiments show that shallow levels are the most beneficial for learned modeling, as they exhibit higher branching uncertainty and influence a large portion of the hierarchy, whereas deeper levels increasingly approach a near-unary regime in which deterministic coding becomes more attractive. This key observation suggests that future real-time learned codecs may benefit from explicitly optimizing ``saved bits per unit compute'' rather than uniformly applying neural inference across all levels. From this perspective, \method{} is an example of compute-aware hybrid entropy modeling for edge deployment.

The heterogeneous-platform experiment further highlights another practical issue: lossless learned entropy coding requires the encoder and decoder to reconstruct exactly the same discrete probability tables. In homogeneous server environments this requirement is often overlooked, but in edge-to-host deployment it becomes a hard system constraint. Our bit-exact entropy coding addresses this issue by moving the entropy synchronization boundary from floating-point probabilities to consistent integer CDF construction.

While \method{} demonstrates significant advantages in runtime, runtime stability, energy efficiency, and cross-platform deployment, several limitations remain and motivate future work. First, although \method{} achieves a favorable system-level trade-off in our evaluated mobile streaming setting, it still introduces a modest increase in compressed data size relative to RENO. A more desirable direction would be to preserve the runtime advantage while further reducing or eliminating this size overhead. 
Second, our current codec focuses on geometry-only compression of independently coded frames after quantization. Temporal redundancy across frames is not yet exploited, and therefore additional gains may be possible in dynamic streaming settings. Third, the current E2E evaluation uses a queueing model with bandwidth caps and isolates runtime effects from protocol-layer dynamics such as packet loss, retransmission, congestion control, and rendering delay. This choice is intended to isolate the encode/decode procedure and compare how codec computation and payload size affect E2E streaming behavior, while a real-world network deployment of the current codec would provide a more comprehensive system-level validation.

Several directions follow naturally from these limitations. A first direction is temporal extension. The shallow--deep split could be combined with inter-frame correlations, motion-compensated occupancy prediction, or incremental update mechanisms so that both spatial hierarchy and temporal correlation are exploited. A second direction is adaptive split control. In the current design, the split depth is fixed offline. A more advanced system could adapt the shallow/deep boundary online according to scene density, available bandwidth, or target latency, thereby turning the codec into a runtime-adaptive streaming primitive. A third direction is joint codec--network co-design. Since our experiments show that queue buildup depends on both data size and computation time, future work could couple compression mode selection with network scheduling, bandwidth estimation, or cross-layer transport control for more stable E2E performance.

Another promising extension is broader cross-platform reproducibility. While the proposed bit-exact entropy coding resolves the heterogeneous decoding issue observed in our setup, future work could formalize this procedure more generally, including standardized integer probability construction, reproducible inference kernels, and deployment guidelines across different accelerators and software stacks. This would be valuable for mobile settings in which sensing, transmission, and consumption may occur on different hardware generations.

Finally, an important next step is task-aware evaluation. In many mobile deployment scenarios, the value of a 3D stream is not measured only by bitrate or latency, but by how well it supports downstream perception, remote interaction, or human-in-the-loop control. Extending the evaluation from codec metrics to task metrics (e.g. operator response quality, reconstruction usability, or downstream perception) would provide a more complete understanding of how edge-friendly geometry compression should be designed for real deployments.
\section{Conclusion}
\label{sec:conclusion}

This paper presents \method{}, a hierarchical compute-aware point cloud compression codec for mobile deployment. \method{} leverages learned coding for shallow sparse occupancy levels, where occupancy uncertainty is higher, and tackles the heavy deep levels with fast deterministic coding. We implement a prototype of \method{} on an NVIDIA Jetson Orin Nano and show that the proposed hybrid shallow--deep design achieves a more favorable compute--rate trade-off in mobile streaming settings.

Experimental results demonstrate that \method{} achieves approximately $3.2\times$--$4.3\times$ faster encoding and $3.5\times$--$5.1\times$ faster decoding than RENO, while also reducing runtime variability. Under bandwidth-limited streaming, it delivers lower and more stable E2E latency. Furthermore, it reduces total energy consumption by up to $5.1\times$ on the Jetson Orin Nano edge device.  Moreover, the proposed bit-exact entropy coding completely eliminates cross-platform decoding failure observed in the learned baseline codec. Overall, these findings demonstrate that compute-aware hybrid codec design is a viable pathway toward deployable real-time 3D streaming on resource-constrained platforms.

\bibliographystyle{IEEEtran}
\bibliography{refs}

\begin{IEEEbiography}[{\includegraphics[width=0.95in,clip,keepaspectratio]{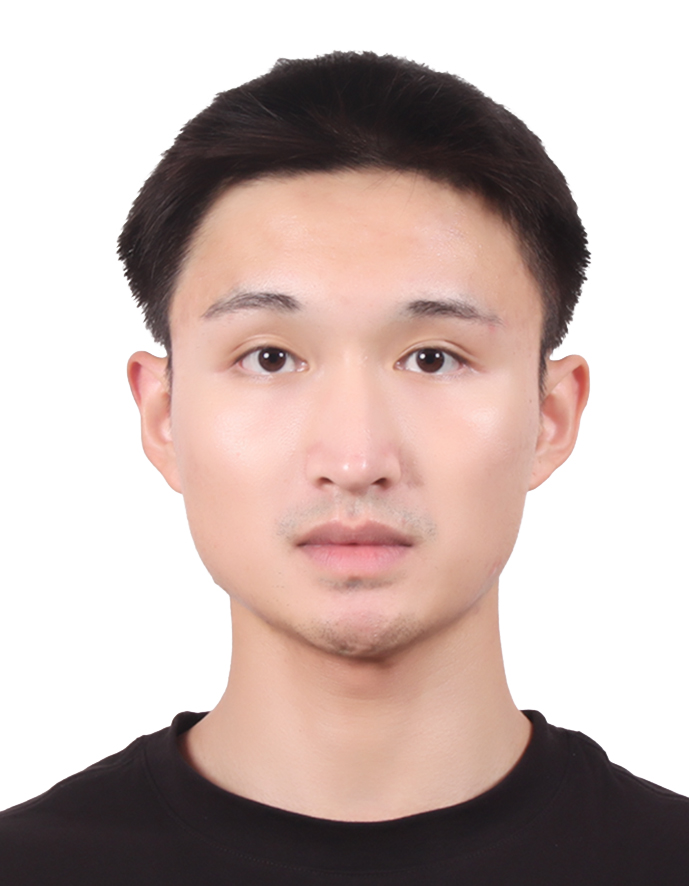}}]{Yuchen Gao} is a Marie Skłodowska-Curie Actions Doctoral Candidate (MSCA DC) with the Department of Electrical and Computer Engineering, Aarhus University, Aarhus, Denmark. His research interests include point cloud compression, low-latency 3D streaming, edge-intelligent 3D systems and immersive 3D communication for Tactile Internet applications.
\end{IEEEbiography}
 
\begin{IEEEbiography}[{\includegraphics[width=0.95in,clip,keepaspectratio]{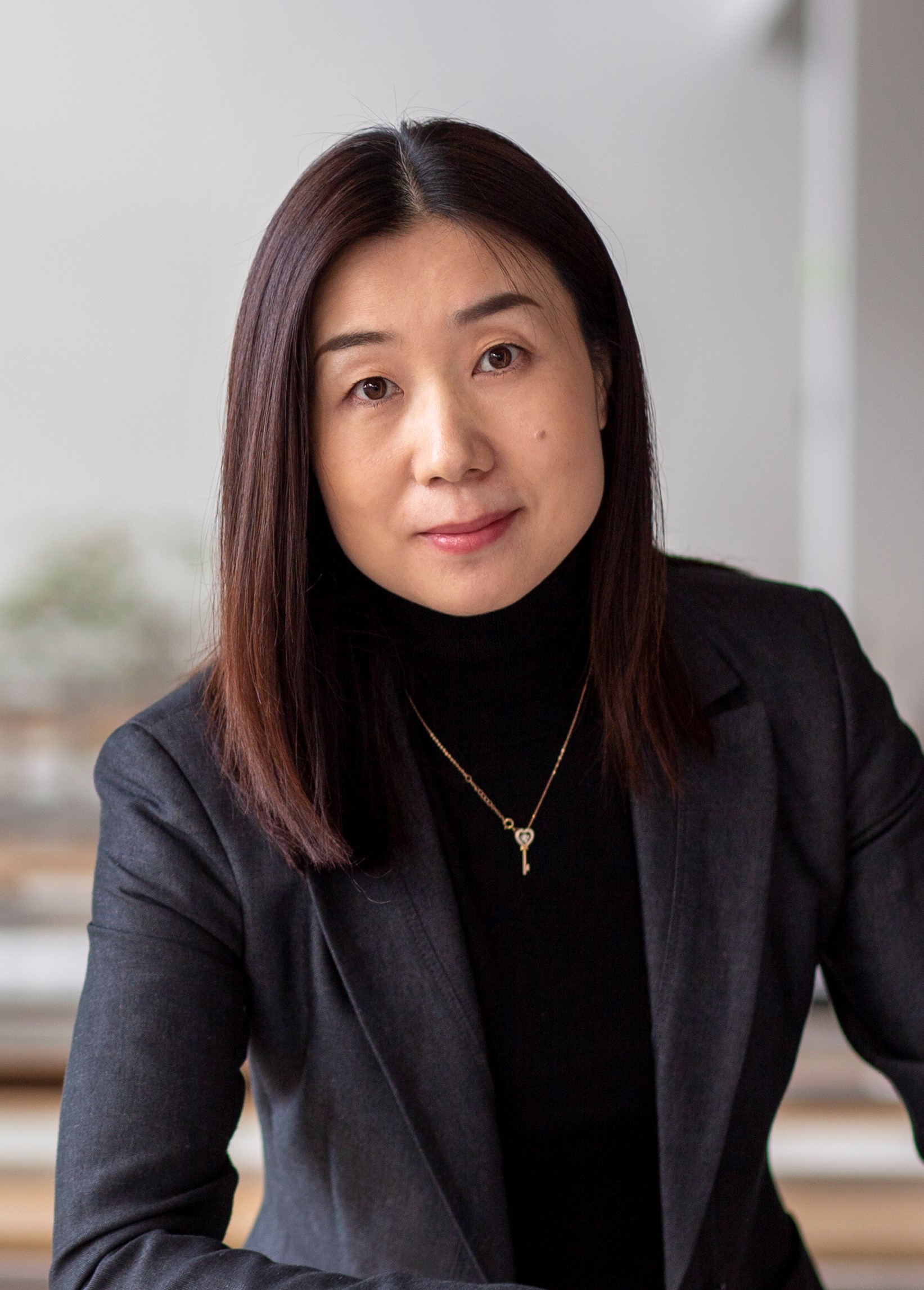}}]{Qi Zhang} (SM’21) is Professor with the Department of Electrical and Computer Engineering, Aarhus University, Aarhus, Denmark. She is currently leading the Internet of Things research area of AU Research Centre for Digitalisation, Big Data and Data Analytics (DIGIT). Her research interests include Edge Intelligence, 6G, Tactile Internet, Goal-oriented Semantic Communication and Internet of Things. 
She is a recipient of three Danish Independent Research grants: AgilE-IoT, Light-IoT and eTouch. She is the project coordinator and a Principle Investigator of Horizon Europe MSCA Doctoral Networks TOAST (Touch-enabled Tactile Internet Training Network and Open Source Testbed). She previously served as an Associate Editor of EURASIP Journal on Wireless Communications and Networking. She has (co-)authored more than 140+ publications in high impact journals and flagship conferences. 
 
\end{IEEEbiography}

\vfill

\end{document}